\documentclass[english,prd,twocolumn,superscriptaddress,floatfix,nofootinbib,preprintnumbers,eqsecnum]{revtex4-1}
\usepackage{graphicx}
\usepackage{amsmath,amsthm,amssymb}
\usepackage{bm}

\usepackage{amsfonts}
\usepackage{dcolumn}
\usepackage{hyperref}

\def\be{\begin{equation}}
\def\ee{\end{equation}}
\def\ba{\begin{eqnarray}}
\def\ea{\end{eqnarray}}
\def\bs{\begin{subequations}}
\def\es{\end{subequations}}

\usepackage{color}

\newcommand{\s}{{\cal S}}

\makeatother

\usepackage{babel}
\makeatother

\begin{document}

\title{Cosmological disformal transformations to the Einstein frame \\ and
gravitational couplings 
with matter perturbations}

\author{Shinji Tsujikawa}

\affiliation{Department of Physics, Faculty of Science, 
Tokyo University of Science,
1-3, Kagurazaka, Shinjuku-ku, Tokyo 162-8601, Japan}

\begin{abstract}

The disformal transformation of metric $g_{\mu \nu} \to 
\Omega^2 (\phi)g_{\mu \nu}+\Gamma(\phi,X) \partial_{\mu}\phi 
\partial_{\nu}\phi$, where $\phi$ is a scalar field with the kinetic 
energy $X= \partial_{\mu}\phi \partial^{\mu}\phi/2$, preserves 
the Lagrangian structure of Gleyzes-Langlois-Piazza-Vernizzi (GLPV) 
theories (which is the minimum extension of Horndeski theories).
In the presence of matter, this transformation gives 
rise to a kinetic-type coupling between the scalar field $\phi$ 
and matter. We consider the Einstein frame in which 
the second-order action of tensor perturbations on the isotropic 
cosmological background is of the same 
form as that in General Relativity and study the role 
of couplings at the levels of both
background and linear perturbations.
We show that the effective gravitational potential felt by matter 
perturbations in the Einstein frame can be conveniently expressed 
in terms of the sum of a General Relativistic contribution and 
couplings induced by the modification of gravity.
For the theories in which the transformed action belongs to a class 
of Horndeski theories, there is no anisotropic stress 
between two gravitational potentials in the Einstein frame 
due to a gravitational de-mixing.
We propose a concrete dark energy model encompassing 
Brans-Dicke theories as well as theories with the tensor propagation 
speed $c_{\rm t}$ different from 1. 
We clarify the correspondence between physical quantities in the 
Jordan/Einstein frames and study the evolution of gravitational 
potentials and matter perturbations from the matter-dominated epoch 
to today in both analytic and numerical approaches.

\end{abstract}

\date{\today}


\maketitle

\section{Introduction}

The large-distance modification of gravity has been under 
active study in connection to the dark energy 
problem \cite{CST,moreview}. 
Modifications of the Einstein-Hilbert term $R/(16\pi G)$ in the 
Lagrangian of General Relativity (GR), where $R$ is the Ricci 
scalar and $G$ is the Newton gravitational constant, generally give rise 
to a radiative scalar degree of freedom $\phi$ \cite{frreview,frreview2}. 
Provided that the fifth force mediated by this new degree 
of freedom is suppressed in the solar system through 
Vainshtein \cite{Vain} or chameleon \cite{Chame} mechanisms, 
the same scalar field can potentially be the source for the 
late-time cosmic acceleration. 

Horndeski theories \cite{Horndeski} are known as the most general 
scalar-tensor theories with one scalar degree of freedom whose 
equations of motion are kept up to second-order in time and 
spatial derivatives (see also Refs.~\cite{Deff}).
Many of dark energy models proposed in the literature-- such as 
$f(R)$ gravity \cite{fRearly,fRviable}, Brans-Dicke (BD) theory \cite{BDpapers,Uddin}, 
kinetic braidings \cite{braid}, and Galileons \cite{Nicolis,cova}-- 
belong to a sub-class of Horndeski  theories. 
In the presence of additional matter, the authors in Ref.~\cite{Koba} 
derived linear perturbation equations of motion on the flat 
Friedmann-Lema\^{i}tre-Robertson-Walker (FLRW) background
to confront dark energy models in the framework of Horndeski theories 
with the observations of large-scale structures, weak lensing, and 
Cosmic Microwave Background (CMB) 
(see also Refs.~\cite{DTHor,Kimura,Kunz,Okada,Bellini,Perenon}).

In BD theories, a scalar degree of freedom $\phi$ is coupled to the 
Ricci scalar $R$ of the form $\phi R$ \cite{Brans}. 
The frame in which matter fields are minimally coupled
to the metric is dubbed the Jordan frame (JF). 
The standard interpretation of measurements is usually
performed in this frame.
In the JF of BD theories, the scalar field $\phi$ mediates 
a fifth force with matter through its gravitational 
interaction with the metric. 
This interaction can be clearly seen in the Einstein frame (EF) 
where the Lagrangian is described by the Einstein-Hilbert term 
plus a canonical scalar field \cite{Fujii,Maeda,Faraoni}. 
In the EF, matter fields feel a metric $g_{\mu \nu}$ conformally 
related to the EF metric $\hat{g}_{\mu \nu}$ of the form 
$\hat{g}_{\mu \nu}=\Omega^2(\phi)g_{\mu \nu}$, where 
$\Omega(\phi)$ is a conformal factor that depends 
on $\phi$ \cite{book,frreview2,Chame}.

For the theories in which field derivatives are coupled to 
the metric, the conformal transformation can be 
extended to a more general mapping of the metric-- dubbed 
the disformal transformation \cite{Bekenstein}. 
In fact, the structure of the Lagrangian in Horndeski theories is 
preserved under the so-called disformal transformation 
$\hat{g}_{\mu \nu}=\Omega^2(\phi)g_{\mu \nu}+
\Gamma (\phi)\partial_{\mu}\phi \partial_{\nu}\phi$, 
where $\Omega$ and $\Gamma$ are functions of 
$\phi$ \cite{Liberati,Garcia}. 
In the presence of matter fields, the disformal transformation
helps us to understand the physical content of mixing 
between $\phi$ and matter \cite{Zum,Koivisto,Bruck,Brax,Minami,Sakstein,Man,Miguel}.

Gleyzes, Langlois, Piazza, and Vernizzi (GLPV) \cite{GLPV} 
proposed a generalized version of  Horndeski theories by 
expressing the Horndeski Lagrangian 
in terms of the Arnowitt-Deser-Misner (ADM) decomposition of 
space-time \cite{ADM} with the choice of the unitary gauge. 
This generally generates derivatives higher than second order, 
but there is only one radiative scalar degree of 
freedom on the FLRW background \cite{Lin,GaoHa}. 
The Lagrangian structure of GLPV theories is preserved
under the disformal transformation of the form
\be
\hat{g}_{\mu \nu}=\Omega^2(\phi)g_{\mu \nu}+
\Gamma (\phi, X)\partial_{\mu}\phi \partial_{\nu}\phi\,,
\label{diformaltra}
\ee
where $\Gamma$ depends on $\phi$ and its kinetic energy 
$X=\partial_{\mu}\phi \partial^{\mu}\phi/2$ \cite{Gleyzes14}.

In the single field system, it was shown that the invariance of curvature 
perturbations $\zeta$ and tensor perturbations $\gamma_{ij}$ holds 
under the disformal transformation (\ref{diformaltra}) \cite{Tsuji14,Naruko,Motohashi} 
(see also Refs.~\cite{Fakir,Minami,Deru,Dome,Yuan} for related works). 
For appropriate choices of $\Omega$ and $\Gamma$, it is possible 
to transform the action to that in the EF where the second-order action of tensor 
perturbations is of the same form as that 
in GR \cite{Vernizzi,Tsuji14,DEGW}. 
This property is useful for the computation of primordial scalar and 
tensor power spectra generated during inflation \cite{Tsuji14}. 
Since the leading-order tensor power spectrum in GLPV theories 
is proportional to the Hubble parameter squared $\hat{H}^2$ in the EF, 
the detection of primordial gravitational waves can determine 
the energy scale $\hat{H}$ during inflation.

In GLPV theories, even though matter is minimally coupled to gravity 
in the JF, there is a mixture between the propagation speeds of 
the scalar field $\phi$ and matter \cite{Gergely,GLPV}.
This comes from a kinetic-type mixing associated with the 
presence of higher-order derivatives beyond the Horndeski domain, 
which is weighed by a parameter $\alpha_{\rm H}$ characterizing 
the deviation from Horndeski theories \cite{GLPV,Kase14,DKT}.
The disformal transformation (\ref{diformaltra}), which contains 
higher-order derivatives, is helpful to understand 
the origin of such a kinetic mixing affecting the scalar and 
matter sound speeds \cite{Gleyzes14}.

If the scalar degree of freedom $\phi$ in Horndeski theories 
is responsible for dark energy, the tensor propagation 
speed $c_{\rm t}$ is typically close to 1 during the 
early cosmological epoch \cite{Tsujikawa15}. 
This is not the case for GLPV theories, in 
which the deviation from $c_{\rm t}=1$ is allowed due 
to the absence of extra conditions Horndeski theories obey.
Recently, dark energy models with constant $c_{\rm t}$ \cite{DKT}
and varying $c_{\rm t}$ \cite{Tsujikawa15} have been proposed 
in the framework of GLPV theories. 
In particular, the latter provides an interesting 
possibility of realizing weak gravity for the perturbations 
relevant to redshift-space distortions \cite{weak,VIPERS}.

In GLPV theories with $c_{\rm t}$ different from 1, 
the disformal transformation (\ref{diformaltra}) to the EF 
should allow us to understand the structure of the matter-scalar 
couplings mentioned above. 
For the models proposed in Refs.~\cite{DKT,Tsujikawa15} 
the anisotropy parameter $\eta=-\Phi/\Psi$ between 
two gravitational potentials $\Psi$ and $\Phi$ deviates from 1 
in the JF, but we will see that it is possible to 
de-mix the gravitational potentials and the scalar field 
in such a way that there is no anisotropic 
stress between $\Psi$ and $\Phi$ in the EF. 
Moreover, we will show that the effective gravitational coupling 
with matter can be well understood in the EF due to the separation of 
a GR-like contribution and modifications arising from $\alpha_{\rm H}$.

In this paper we obtain relations of physical quantities between 
the JF and the EF under the disformal transformation (\ref{diformaltra}) 
and study roles of gravitational couplings with matter at the levels of 
both background and linear perturbations.
A similar prescription was taken in Ref.~\cite{Man} 
with the disformal transformation 
$\hat{g}_{\mu \nu}=\Omega^2(\phi)g_{\mu \nu}+
\Gamma (\phi)\partial_{\mu}\phi \partial_{\nu}\phi$, but 
our treatment is more general in that it is not restricted to 
the transformation between Horndeski theories alone.  
We employ the approach of effective field theory of cosmological 
perturbations \cite{Weinberg,Park,Bloomfield,Battye,Mueller,Gubi,Piazza2,Silve}, 
allowing to encompass GLPV theories as a specific case. 
We propose a new dark energy model in the framework of 
GLPV theories, which accommodates models with 
$c_{\rm t}^2 \neq 1$ as well as models based on BD theories
(which lead to ``coupled quintessence'' 
models \cite{Amendola} in the EF).

Our paper is organized as follows.
In Sec.~\ref{modelsec} we briefly review GLPV theories 
and in Sec.~\ref{disformalsec} we show how the GLPV action 
is transformed under the disformal transformation 
(\ref{diformaltra}) in the presence of matter. 
In Sec.~\ref{transsec} we present linear perturbation equations 
of motion on the flat FLRW background 
in both the JF and the EF. 
In Sec.~\ref{Einsec} we consider the transformation to the EF 
and discuss the matter-scalar coupling in the EF.
In Sec.~\ref{newmodelsec} we propose a new dark energy model 
belonging to GLPV theories and study the correspondence of physical 
quantities between the JF and the EF in detail.
Section \ref{consec} is devoted to conclusions.

\section{GLPV theories in the presence of matter}
\label{modelsec}

GLPV theories \cite{GLPV} are the generalizations of 
Horndeski theories written in terms of ADM 
scalar quantities defined below \cite{building}. 
We begin with the line element 
\ba
\hspace{-0.3cm}
ds^{2} &=&
g_{\mu \nu }dx^{\mu }dx^{\nu} \nonumber \\
\hspace{-0.3cm}
&=& -N^{2}dt^{2}+h_{ij}(dx^{i}+N^{i}dt)(dx^{j}+N^{j}dt)\,,  
\label{ADMmetric}
\ea
where $N$ is the lapse function, $N^i$ is the shift vector, and
$h_{ij}$ is the three-dimensional spatial metric. 
We express the three-dimensional Ricci tensor on
the constant time hyper-surfaces $\Sigma_t$, as 
${\cal R}_{\mu \nu}={}^{(3)}R_{\mu \nu}$.
The extrinsic curvature is defined by 
$K_{\mu \nu}=h^{\lambda}_{\mu} n_{\nu;\lambda}$, 
where $n_{\mu}=(-N,0,0,0)$ is a unit vector 
orthogonal to $\Sigma_t$. 
We introduce a number of geometric scalar quantities, as
\ba
& &
K \equiv {K^{\mu}}_{\mu}\,,\quad
\s \equiv K_{\mu \nu} K^{\mu \nu}\,,\nonumber \\
& &
{\cal R} \equiv
{{\cal R}^{\mu}}_{\mu}\,,\quad
{\cal U} \equiv {\cal R}_{\mu \nu} K^{\mu \nu}\,. 
\label{ADMscalar}
\ea

Horndeski theories, which have one scalar degree 
of freedom $\phi$, can be reformulated by using 
the above geometric scalars with the choice 
of unitary gauge
\be
\phi=\phi(t)\,,
\label{uni}
\ee
under which $\phi$ depends on the cosmic time $t$ alone.
On the flat FLRW background with the scale factor $a(t)$, 
the Lagrangian of Horndeski theories can be expressed 
in the form \cite{building}
\ba
L  &=&
A_2(N,t)+A_3(N,t)K \nonumber \\
& &
+A_4(N,t) (K^2-{\cal S}) +B_4(N,t){\cal R} 
\nonumber \\
& &
+A_5(N,t) K_3 
+B_5(N,t) \left( {\cal U}-\frac12 K {\cal R} \right)\,,
\label{LGLPV}
\ea
where $K_3 \equiv K^3-3KK_{\mu \nu}K^{\mu \nu}
+2K_{\mu \nu}K^{\mu \lambda}{K^{\nu}}_{\lambda}$, 
and $A_i$, $B_i$ are functions of $N$ and $t$ 
satisfying the two conditions \cite{GLPV}
\be
A_4 = 2XB_{4,X}-B_4\,,\qquad
A_5 =-\frac13 XB_{5,X}\,,
\label{ABcon}
\ee
where $B_{i,X} \equiv \partial B_i/\partial X$ with
$X \equiv g^{\mu \nu} \partial_{\mu}\phi \partial_{\nu}\phi$. 
In the unitary gauge we have $X=-\dot{\phi}^2(t)/N^2$, 
where a dot represents a derivative with respect to $t$. 
Hence the dependence on $N$ and $t$ translates to 
that on $X$ and $\phi$.

Violation of the conditions (\ref{ABcon}) can generally give 
rise to derivatives higher than second order, but it was shown 
in Refs.~\cite{GLPV,Lin,GaoHa} that 
there is only one scalar propagating degree of 
freedom on the flat FLRW background. 
GLPV theories are described by the Lagrangian 
(\ref{LGLPV}) without having the two conditions (\ref{ABcon}).
In this paper we focus on GLPV theories in the presence of 
a matter field $\Psi_m$ described by the Lagrangian $L_m$.
Then, we consider the following action 
\ba
S &=&
\int d^4 x \sqrt{-g}\,L(N,K, {\cal S}, {\cal R}, {\cal U};t) 
\nonumber \\
& &+\int d^4 x \sqrt{-g}\,L_m (g_{\mu \nu},\Psi_m)\,,
\label{action}
\ea
where $g$ is the determinant of metric $g_{\mu \nu}$, and 
$L$ is given by Eq.~(\ref{LGLPV}). 
The matter field $\Psi_m$ is assumed to be 
a barotropic perfect fluid, which 
can be modeled by a k-essence Lagrangian $P(Y)$ 
depending on the kinetic term 
$Y=g^{\mu \nu}\partial_{\mu}\chi\partial_{\nu}\chi$ 
of a scalar field $\chi$ \cite{Arroja,Kase14,DKT}.
The term $K_3$ in Eq.~(\ref{LGLPV}) can be expressed 
as $K_3=3H (2H^2-2KH+K^2-{\cal S})$
up to second order in the perturbations \cite{building}, 
where $H$ is the Hubble parameter defined later 
in Eq.~(\ref{Hubbledef}). 
Hence the Lagrangian (\ref{LGLPV}) of GLPV theories 
depends on $N, K, {\cal S}, {\cal R}, {\cal U}$, and $t$
up to linear order in the perturbations.

We assume that, in the JF, the matter field $\Psi_m$ 
is minimally coupled to the metric $g_{\mu \nu}$.
The matter energy-momentum tensor following from 
$L_m$ is given by 
\be
T^{\mu \nu}=\frac{2}{\sqrt{-g}}
\frac{\delta (\sqrt{-g}\,L_m)}{\delta g_{\mu \nu}}\,.
\label{Tmunu}
\ee
We can derive the background and linear perturbation equations of 
motion by varying the action (\ref{action}) up to first and second orders 
in the perturbations, respectively \cite{building,Gergely,Kase2}. 

\section{Disformal transformations}
\label{disformalsec}

In this section we discuss how background/perturbed quantities 
and the action (\ref{action}) are mapped
under the disformal transformation (\ref{diformaltra}). 

\subsection{Transformation of background and perturbed quantities}

In the unitary gauge (\ref{uni}), the line element 
$d\hat{s}^2=\hat{g}_{\mu \nu}dx^{\mu}dx^{\nu}$
in the transformed frame reads \cite{Minami,Tsuji14}
\be
d\hat{s}^2=-\hat{N}^2dt^2+\hat{h}_{ij}
(dx^{i}+N^{i}dt)(dx^{j}+N^{j}dt)\,,
\ee
where 
\be
\hat{N} = N\alpha\,,\qquad 
\hat{h}_{ij} = \Omega^2 h_{ij}\,,
\label{hath}
\ee
with 
\be
\alpha \equiv 
\sqrt{\Omega^2+\Gamma X}\,.
\label{aldef}
\ee

In the JF, let us consider the linearly perturbed line element 
on the flat FLRW background,
\ba
\hspace{-0.2cm}
ds^2 &=& 
-(\bar{N}^2+2A)dt^2+2 \psi_{|i} dt dx^i \nonumber \\
\hspace{-0.2cm}
& &+a^2(t) [ (1+2\zeta) \delta_{ij} 
+2E_{|ij} +\gamma_{ij} ]dx^i dx^j\,,
\label{permet}
\ea
where $\bar{N}$ is the background value of the lapse; 
 $A, \psi, \zeta, E$ are the scalar metric perturbations; 
$\gamma_{ij}$ is the tensor perturbation, 
and the lower index ``${}_{|i}$'' denotes the covariant 
derivative with respect to the three-dimensional metric $h_{ij}$.
Comparing Eq.~(\ref{ADMmetric}) with Eq.~(\ref{permet}), 
we have the relations $2A=N^2-h_{ij}N^i N^j-\bar{N}^2$, 
$\psi_{|i}=h_{ij}N^j$, and 
\be
h_{ij}=a^2(t) \left[ (1+2\zeta) \delta_{ij}
+2E_{|ij}+\gamma_{ij} \right]\,.
\label{hij}
\ee
Introducing the scale factor $\hat{a}(t)$ in the 
transformed frame, as
\be
\hat{a}(t)=\Omega(t) a(t)\,,
\label{hata}
\ee
the three-dimensional metric $\hat{h}_{ij}$ 
in Eq.~(\ref{hath}) reduces to
\be
\hat{h}_{ij}=\hat{a}^2(t) \left[ (1+2\hat{\zeta}) \delta_{ij}
+2\hat{E}_{|ij}+\hat{\gamma}_{ij} \right]\,,
\label{hij2}
\ee
where
\be
\hat{\zeta}=\zeta\,,\qquad 
\hat{E}=E\,,\qquad
\hat{\gamma}_{ij} =\gamma_{ij}\,.
\label{zetaE}
\ee
In what follows we use an over-hat for quantities in the transformed 
frame. {}From Eq.~(\ref{zetaE}), the perturbations $\zeta$, $E$, and $\gamma_{ij}$ 
are invariant under the disformal transformation (\ref{diformaltra}) \cite{Tsuji14}.
{}From Eq.~(\ref{hath}) and the relation $\hat{N}^i=N^i$ 
we also obtain 
\be
\hat{\delta N}=\frac{1}{\bar{\beta}} \delta N\,,\qquad
\hat{\psi}=\Omega^2 \psi\,,
\label{Npsieq}
\ee
where an over-bar represents background quantities, and 
\be
\beta \equiv \frac{\sqrt{\Omega^2+\Gamma X}}
{\Omega^2-X^2\Gamma_{,X}}\,.
\label{betadef}
\ee
Since $\phi=\phi(t)$ and $X=-\dot{\phi}^2(t)/N^2$ in the unitary 
gauge, the quantities $\Gamma$ and $X$ in Eq.~(\ref{betadef}) 
contain the information of perturbations through the lapse function $N$.

\subsection{Transformation of the action}

The disformal transformation of the action $S_g=\int d^4x \sqrt{-g}\,L$, 
where $L$ is the Lagrangian (\ref{LGLPV}) of GLPV theories, was 
already discussed in Refs.~\cite{Gleyzes14,Tsuji14}.
First of all, the volume element $\sqrt{-g}$ transforms as
\be
\sqrt{-\hat{g}}=\sqrt{-g}\,\Omega^3 \alpha\,.
\label{gtra}
\ee
In the unitary gauge, the intrinsic and extrinsic curvatures 
obey the transformation laws
\be
\hat{{\cal R}}_{ij}={\cal R}_{ij}\,,\qquad
\hat{K}_{ij}=\frac{\Omega^2}{\alpha} 
\left( K_{ij}+\frac{\omega}{N}h_{ij} \right)\,,
\label{RKtra}
\ee
where 
\be
\omega \equiv \frac{\dot{\Omega}}{\Omega}\,.
\ee
The action in the transformed frame 
$S_g=\int d^4 x \sqrt{-\hat{g}}\hat{L}$ preserves 
the structure of original GLPV action, such that 
\ba
\hspace{-0.3cm}
\hat{L} &=&
\hat{A}_2(\hat{N},t)+\hat{A}_3(\hat{N},t)\hat{K}
\nonumber \\
\hspace{-0.3cm}
&&
+\hat{A}_4(\hat{N},t) (\hat{K}^2-\hat{\cal S})
+\hat{B}_4(\hat{N},t)\hat{{\cal R}}
\nonumber \\
\hspace{-0.3cm}
&&
+\hat{A}_5(\hat{N},t) \hat{K}_3
+\hat{B}_5(\hat{N},t) \left( \hat{\cal U}-\frac12
\hat{K} \hat{\cal R} \right),
\label{LH2}
\ea
where
\ba
\hspace{-0.5cm}
\hat{A}_2 &=&
\frac{1}{\Omega^3 \alpha} \left( A_2-\frac{3\omega}{N}A_3
+\frac{6\omega^2}{N^2}A_4-\frac{6\omega^3}{N^3}A_5 \right),
\label{hatA2}\\
\hspace{-0.5cm}
\hat{A}_3 &=&
\frac{1}{\Omega^3}  \left( A_3-\frac{4\omega}{N}A_4
+\frac{6\omega^2}{N^2}A_5 \right),
\label{hatA3}\\
\hspace{-0.5cm}
\hat{A}_4 &=&
\frac{\alpha}{\Omega^3} \left( A_4-\frac{3\omega}{N}A_5 \right),
\label{hatA4} \\
\hspace{-0.5cm}
\hat{B}_4 &=&
\frac{1}{\Omega \alpha} \left( B_4+\frac{\omega}{2N}B_5 \right),
\label{hatB4}\\
\hspace{-0.5cm}
\hat{A}_5 &=&
\frac{\alpha^2}{\Omega^3}A_5,
\label{hatA5} \\
\hspace{-0.5cm}
\hat{B}_5 &=&
\frac{1}{\Omega}B_5\,.
\label{hatB5}
\label{cotra}
\ea

The matter action in the transformed frame is given by 
$S_m=\int d^4x \sqrt{-\hat{g}}\,\hat{L}_m$, where 
\be
\hat{L}_m=\frac{1}{\Omega^3 \alpha}L_m\,,
\label{hatLm}
\ee
where $L_m$ depends on the JF metric 
$g_{\mu \nu}$ and the matter field $\Psi_m$.
By expressing $g_{\mu \nu}$ in terms of the metric 
$\hat{g}_{\mu \nu}$ in the transformed frame from 
Eq.~(\ref{diformaltra}), it contains the contribution of 
$\phi$ and its derivative. 
Hence the scalar field $\phi$ is (kinetically) 
coupled to matter in the transformed frame.

{}From Eq.~(\ref{Tmunu}) the transformation law 
of $T_{\mu \nu}$ is given by 
\be
T^{\mu \nu}=\frac{\sqrt{-\hat{g}}}{\sqrt{-g}} 
\frac{\delta \hat{g}_{\gamma \rho}}{\delta g_{\mu \nu}} 
\hat{T}^{\gamma \rho}\,.
\ee
On using Eqs.~(\ref{diformaltra}) and (\ref{gtra}), 
it follows that \cite{Garcia}
\be
T^{\mu \nu}=\hat{T}^{\gamma \rho} \Omega^3 \alpha 
\left( \Omega^2 \delta^{\mu}_{\gamma} \delta^{\nu}_{\rho}
-\Gamma_{,X} \partial^{\mu}\phi \partial^{\nu}\phi
\partial_{\gamma} \phi  \partial_{\rho} \phi \right)\,.
\label{Tmunu2}
\ee
The transformed metric with upper indices is given by 
\be
\hat{g}^{\mu \nu}=\frac{1}{\Omega^2} 
\left( g^{\mu \nu} -\frac{\Gamma}{\alpha^2} 
\partial^{\mu} \phi \partial^{\nu} \phi \right)\,.
\label{gmunuin}
\ee
From Eqs.~(\ref{Tmunu2}) and (\ref{gmunuin}),  
the mixed energy-momentum tensor obeys 
the transformation law:
\ba
T^{\mu}_{\lambda} &=& \hat{T}^{\gamma}_{\sigma}\,\Omega\,\alpha
\biggl[ \Omega^2 \delta^{\mu}_{\gamma} \left( \delta^{\sigma}_{\lambda}
-\frac{\Gamma}{\alpha^2} \partial^{\sigma} \phi \partial_{\lambda} \phi 
\right) \nonumber \\
& &
\qquad \quad 
-\frac{\Omega^2 \Gamma_{,X}}{\alpha^2} 
\partial_{\gamma} \phi \partial^{\sigma} \phi
\partial_{\lambda} \phi \partial^{\mu} \phi \biggr]\,.
\ea
For the choice of the unitary gauge (\ref{uni}), we obtain the relations
\be
T^0_0=\hat{T}^0_0 \frac{\Omega^3}{\beta}\,,\quad
T^0_i=\hat{T}^0_i\,\Omega^3 \alpha \,,\quad
T^i_j=\hat{T}^i_j\,\Omega^3 \alpha \,.
\ee
We decompose the energy-momentum tensor into the background and 
perturbed parts, as $T^{0}_0=-\rho-\delta \rho$, $T^0_i=\partial_i\delta q$, and 
$T^i_j=(P+\delta P)\delta^i_j$, where $T^{0}_i$ is a perturbed quantity itself. 
The background energy density $\rho$ and the pressure $P$ are
subject to the transformations
\be
\hat{\rho}=\frac{\bar{\beta}}{\Omega^3}\rho\,,\qquad
\hat{P}=\frac{1}{\Omega^3 \bar{\alpha}} P\,.
\label{backre}
\ee
For the linear perturbations, we have 
\ba
\hat{\delta \rho} &=&
\frac{\bar{\beta}}{\Omega^3} \delta \rho+
\frac{\nu}{\Omega^3}\rho\,\delta N\,,\\
\hat{\delta q} &=& \frac{1}{\Omega^3 \bar{\alpha}}
\delta q\,,\label{deltaq}\\
\hat{\delta P} &=& \frac{1}{\Omega^3 \bar{\alpha}}
\left( \delta P-\mu P\,\delta N \right)\,,
\label{perre}
\ea
where 
\be
\mu \equiv \frac{1}{\alpha} \frac{\partial \alpha}
{\partial N}\biggr|_{N=\bar{N}}\,,\qquad
\nu \equiv \frac{\partial \beta}{\partial N} \biggr|_{N=\bar{N}}\,.
\label{numu}
\ee
The quantity $\mu$ is related to $\bar{\alpha}$ and 
$\bar{\beta}$, as
\be
\mu \bar{N}=\frac{1}{\bar{\alpha}\bar{\beta}}-1\,.
\label{mure}
\ee

In summary, the action in the transformed frame reads
\ba
S &=&
\int d^4 x \sqrt{-\hat{g}}\,\hat{L}(\hat{N},\hat{K}, \hat{{\cal S}}, 
\hat{{\cal R}}, \hat{{\cal U}};t) 
\nonumber \\
& &+\int d^4 x \sqrt{-\hat{g}}\,\hat{L}_m 
(\hat{g}_{\mu \nu}(\phi,\partial_{\mu}\phi),
\Psi_m)\,,
\label{actiontra}
\ea
where $\hat{L}$ and $\hat{L}_m$ are given, respectively, by 
Eqs.~(\ref{LH2}) and (\ref{hatLm}). 

\section{Equations of motion in the JF and the transformed frame}
\label{transsec}

In this section we present linear perturbation equations of motion on 
the flat FLRW background for the theories described 
by the action (\ref{action}). We then study how they are transformed 
under the disformal transformation (\ref{diformaltra}).

\subsection{Equations of motion in the JF}

The equations of motion in the JF were already derived 
in Refs.~\cite{Gleyzes14,Gergely,Kase14,DKT} 
for the action (\ref{action}). 
We consider the background line element 
$ds^2=-\bar{N}^2dt^2+a^2(t) \delta_{ij}dx^i dx^j$ without 
setting $\bar{N}=1$. Defining the Hubble parameter
\be
H \equiv \frac{\dot{a}}{\bar{N}a}\,,
\label{Hubbledef}
\ee
the background values of extrinsic and intrinsic 
curvatures are given, respectively, by 
\be
\bar{K}_{\mu \nu}=H\bar{h}_{\mu \nu}\,,\qquad
\bar{{\cal R}}_{\mu \nu}=0\,,
\ee
and hence $\bar{K}=3H$,  $\bar{\cal S}=3H^{2}$, and 
$\bar{{\cal R}}=\bar{\cal U}=0$.

The background and perturbation equations of motion follow from 
first-order and second-order Lagrangians, respectively, 
derived by expanding the action (\ref{action}) up to 
quadratic order in perturbations. 
The perturbations of $N,K,{\cal S}$ are given by 
$\delta N=N-\bar{N}$, $\delta K=K-3H$, 
and $\delta {\cal S}=2H\delta K+\delta
K_{\nu }^{\mu} \delta K_{\mu}^{\nu}$. 
We write the intrinsic curvature as 
${\cal R}=\delta_1 {\cal R}+\delta_2 {\cal R}$, where 
$\delta_1 {\cal R}$ and $\delta_2 {\cal R}$ are first-order 
and second-order perturbations, respectively. 
For the perturbation ${\cal U}$, we have the 
relation $\int d^4x \sqrt{-g}\,\lambda(t) \,{\cal U}
=\int d^4x \sqrt{-g} [\lambda(t) {\cal R} K/2+
\dot{\lambda}(t){\cal R}/(2N)]$ up to a boundary term, 
where $\lambda(t)$ is an arbitrary function 
with respect to $t$ \cite{building}.

We consider the perturbed metric (\ref{permet}) with 
the choice of unitary gauge 
\be
\delta \phi=0\,,\qquad E=0\,,
\ee
under which the temporal and spatial coordinate transformation 
vectors are fixed, respectively.

\subsubsection{Background equations}

Expanding the gravitational action $S_g=\int d^4x \sqrt{-g}\,L$ up to 
first order in the scalar perturbations, 
it follows that \cite{Gleyzes14}
\ba
\delta S_g
&=& \int d^4 x \biggl[
a^3 \left( \bar{L}+\bar{N}L_{,N}-3H {\cal F} \right) \delta N 
\nonumber \\
& &~~~~~~~~~
+3a^2\bar{N} \left( \bar{L}-\frac{\dot{\cal F}}{\bar{N}}
-3H{\cal F} \right)
\delta a \biggr]\,,
\label{Sgexpan}
\ea
where
\be
{\cal F} \equiv L_{,K}+2H L_{,{\cal S}}\,,
\ee
and we dropped a boundary term irrelevant to the dynamics. 
Here and in the following, the coefficients in front of perturbed 
quantities [such as those in front of $\delta N$ and $\delta a$ in 
Eq.~(\ref{Sgexpan})]
should be evaluated on the background.

Variation of the matter energy-momentum 
tensor $\delta S_m=\int d^4 x\sqrt{-g}\,T^{\mu \nu} 
\delta g_{\mu \nu}/2$ reads
\be
\delta S_m=\int d^4x \left( -a^3 \rho\,\delta N
+3a^2 \bar{N}P \delta a \right)\,.
\ee
{}From the variational principle $\delta S_g+\delta S_m=0$, 
we obtain the background equations of motion 
\ba
& &
\bar{L}+\bar{N}L_{,N}-3H {\cal F}=\rho\,,\label{back1} \\
& &
\bar{L}-\frac{\dot{\cal F}}{\bar{N}}-3H{\cal F}=-P\,.
\label{back2} 
\ea
Since the matter component is not directly coupled to the 
field $\phi$ in the JF, it obeys the standard continuity equation 
\be
\frac{\dot{\rho}}{\bar{N}}+3H (\rho+P)=0\,.
\label{conti}
\ee

\subsubsection{Perturbation equations}

Expanding the action (\ref{action}) with the Lagrangian (\ref{LGLPV}) 
up to second order in scalar perturbations 
and taking the variation with respect to $\delta N$, $\partial^2 \psi$, 
$\zeta$, and the field perturbation $\delta \chi$ associated with the 
matter Lagrangian $P(Y)$, 
the resulting perturbation equations 
of motion in the presence of a barotropic perfect fluid 
are given, respectively, by \cite{Gleyzes14,Gergely,Kase14,DKT}
\ba
& &
\left( 2\bar{N}L_{,N}+{\bar N}^2L_{,NN}-6H\bar{N}{\cal W}
+12L_{,{\cal S}}H^2 \right) 
\frac{\delta N}{\bar{N}} \nonumber \\
& &+\left( 3\dot{\zeta}-\frac{\partial^2 \psi}{a^2} \right){\cal W}
-4(\bar{N}{\cal D}+{\cal E})\frac{\partial^2 \zeta}{a^2}=\delta \rho\,,
\label{pereq1} \\
& &
{\cal W} \frac{\delta N}{\bar{N}}
-\frac{4L_{,{\cal S}}}{\bar{N}^2} \dot{\zeta}=-\delta q\,,
\label{pereq2} \\
& &
\frac{1}{a^3\bar{N}} \frac{d}{dt} \left( a^3 \bar{N} {\cal Y} \right)
+4(\bar{N}{\cal D}+{\cal E}) \frac{\partial^2 \delta N}{a^2\bar{N}}
+\frac{4{\cal E}}{a^2} \partial^2 \zeta \nonumber \\
& &
-3(\rho+P) \frac{\delta N}{\bar{N}}
=3\delta P\,,
\label{pereq3} \\
& &
\frac{\dot{\delta \rho}}{\bar{N}}+3H \left( \delta \rho+\delta P \right)
\nonumber \\
& &
=-(\rho+P) \left( \frac{3\dot{\zeta}}{\bar{N}}
-\frac{\partial^2 \psi}{a^2\bar{N}} \right)
-\bar{N}\frac{\partial^2\delta q}{a^2}\,,
\label{pereq4}
\ea
where 
\ba
{\cal D} &\equiv&
L_{,N\mathcal{R}}-\frac{\dot{L_{,{\cal U}}}}{2\bar{N}^2}
+HL_{,N {\cal U}}\,, \label{calD} \\
{\cal E} &\equiv& L_{,{\cal R}}+\frac{\dot{L}_{,\cal U}}{2\bar{N}}
+\frac32 HL_{,\cal U}\,,\\
{\cal W} &\equiv& 
L_{,KN}+2HL_{,{\cal S}N} +\frac{4HL_{,\cal S}}{\bar{N}} \,,  \\
{\cal Y} &\equiv& \frac{4L_{,{\cal S}}}{\bar{N}^2} 
\frac{\partial^2 \psi}{a^2}
-3\delta q\,.
\label{W}
\ea
The momentum perturbation $\delta q$ obeys 
\be
\frac{1}{\bar{N}} \frac{d}{dt} (\bar{N} \delta q)+3H \bar{N} \delta q
=-(\rho+P)\frac{\delta N}{\bar{N}}-\delta P\,.
\label{deltaqeq}
\ee

Substituting Eq.~(\ref{W}) into Eq.~(\ref{pereq3}) and 
using Eq.~(\ref{deltaqeq}), it follows that
\ba
& &
\left( \frac{\dot{L}_{,\cal S}}{\bar{N}}+HL_{,\cal S}
-\frac{\dot{\bar N}}{{\bar N}^2}L_{,\cal S} \right) 
\frac{\psi}{\bar{N}}+L_{,\cal S} \frac{\dot{\psi}}
{\bar{N}^2} \nonumber \\
& &+(\bar{N}{\cal D}+{\cal E}) \frac{\delta N}{\bar{N}}
+{\cal E} \zeta=0\,,
\label{anieq1}
\ea
where we set the integration constant 0.
We define the gauge-invariant Bardeen potentials \cite{Bardeen}
\be
\Psi \equiv \frac{\delta N}{\bar{N}}+\frac{1}{\bar{N}} 
\frac{d}{dt} \left( \frac{\psi}{\bar{N}} \right)\,,\qquad
\Phi \equiv \zeta+H\frac{\psi}{\bar{N}}\,,
\label{PsiPhi}
\ee
and the anisotropy parameter 
\be
\eta \equiv -\frac{\Phi}{\Psi}\,.
\ee
The effective gravitational potential associated with 
the deviation of light rays in CMB and weak lensing 
observations is given by \cite{Sapone}
\be
\Phi_{\rm eff}=\frac12 (\Psi-\Phi)
=\frac12 (1+\eta)\Psi\,.
\ee

One can write Eq.~(\ref{anieq1}) in the form
\be
\Psi+\Phi=\frac{\dot{L}_{,\cal S}}{\bar{N}H L_{,\cal S}} 
(\zeta-\Phi)-\left( \frac{c_{\rm t}^2}{\bar{N}^2}-1 \right) 
\zeta-\alpha_{\rm H} \frac{\delta N}{\bar{N}}\,,
\label{anire}
\ee
where $c_{\rm t}^2=\bar{N}^2{\cal E}/L_{,{\cal S}}$ 
is the tensor propagation speed squared discussed 
later in Sec.~\ref{Einsec}, and 
\be
\alpha_{\rm H} \equiv \frac{\bar{N}{\cal D}+{\cal E}}
{L_{,\cal S}}-1=\frac{c_{\rm t}^2}{\bar{N}^2}-1
+\frac{\bar{N}{\cal D}}{L_{,\cal S}}\,.
\ee
The parameter $\alpha_{\rm H}$ characterizes the 
deviation from Horndeski theories.
Provided that one of the conditions $\dot{L}_{,\cal S} \neq 0$, 
$c_{\rm t}^2 \neq \bar{N}^2$, and $\alpha_{\rm H} \neq 0$ is 
satisfied, the anisotropic stress does not generally 
vanish ($\eta \neq 1$).

We define the gauge-invariant matter density contrast, as
\be
\delta_m \equiv \delta-3V_m\,,
\ee
where $\delta \equiv \delta \rho/\rho$ and $V_m \equiv \bar{N} H\delta q/\rho$.
Taking the time derivative of Eq.~(\ref{pereq4}) and using Eq.~(\ref{deltaqeq}) 
in Fourier space, we obtain 
\be
\ddot{\delta}_m+2H \dot{\delta}_m
+\frac{k^2}{a^2}\Psi=
-3\ddot{B}-6H\dot{B}\,, 
\label{mattereq}
\ee
where $B \equiv \zeta+V_m$, and $k$ is a coming wave number.
We define the effective gravitational coupling $G_{\rm eff}$, as
\be
\frac{k^2}{a^2}\Psi=-4\pi G_{\rm eff} \rho \delta_m\,.
\ee
The gravitational potential $\Psi$ contains the information of 
gravitational coupling with the scalar field $\phi$. 
The modified gravitational interaction affects the evolution of 
matter perturbations through Eq.~(\ref{mattereq}). 
The evolution of $\Psi$ and $G_{\rm eff}$ is known by 
solving the coupled Eqs.~(\ref{pereq1})-(\ref{pereq4}) 
with Eq.~(\ref{deltaqeq}).

On using Eqs.~(\ref{pereq1}) and (\ref{pereq2}), the second-order action 
of scalar perturbations can be expressed in terms of $\zeta$, 
$\delta \chi$, and its derivatives. 
Assuming that the matter sector does not correspond
to a ghost mode, the scalar ghost is absent under 
the condition \cite{building,Gergely}
\ba
q_{\rm s} &=&
\frac{2L_{,\cal S}(4L_{,\cal S}w_s+3\bar{N}^2{\cal W}^2)}
{\bar{N}^3 {\cal W}^2}>0\,,
\label{qs}
\ea
where $w_s=2\bar{N}L_{,N}+{\bar N}^2L_{,NN}-6H\bar{N}{\cal W}
+12L_{,{\cal S}}H^2$.

In GLPV theories there is a mixing between the scalar propagation 
speed $c_{\rm s}$ and the matter sound speed $c_m$.
For non-relativistic matter characterized by $P= +0$ 
and $\delta P=+0$, we have $c_m^2=+0$ in the small-scale limit, 
while $c_{\rm s}^2$ is given by \cite{Gergely,GLPV,Tsujikawa15}
\be
c_{\rm s}^2=\frac{2\bar{N}}{q_{\rm s}} \left[ 
\frac{\dot{\cal M}}{\bar{N}}+H{\cal M}-{\cal E} 
-\frac{4L_{,\cal S}^2\rho}{\bar{N}^2{\cal W}^2}
(1+2\alpha_{\rm H}) \right]\,,
\label{cs}
\ee
where 
\be
{\cal M} \equiv \frac{4L_{,\cal S}(\bar{N}{\cal D}+{\cal E})}
{\bar{N}{\cal W}}\,.
\ee
This shows that the deviation from Horndeski theories 
($\alpha_{\rm H} \neq 0$) modifies the scalar sound speed. 
We require $c_{\rm s}^2>0$
to avoid Laplacian instabilities. 

\subsection{Equations of motion in the transformed frame}

In the transformed frame described by the action (\ref{actiontra}),
we also derive the background and perturbation equations of motion. 
The cosmic time $\hat{t}$ in the 
transformed frame is related to $t$ in the JF, as 
\be
\hat{t}=\int \hat{\bar{N}} dt\,.
\ee

\subsubsection{Background equations}

Following the same procedure as that in the JF,  
we obtain the background equations of motion 
\ba
& &
\hat{\bar L}+\hat{\bar N} \hat{L}_{,\hat{N}}-3\hat{H}\hat{\cal F}
=\hat{\rho}\,,
\label{backtra1} \\
& & 
\hat{\bar L}-\frac{\dot{\hat{\cal F}}}{\hat{\bar N}}
-3\hat{H}\hat{\cal F}=-\hat{P}\,,
\label{backtra2}
\ea
where $\hat{\cal F} \equiv \hat{L}_{,{\hat{K}}}+
2\hat{H}\hat{L}_{,{\hat{\cal S}}}$ and 
\be
\hat{H} \equiv \frac{\dot{\hat a}}{\hat{\bar N}{\hat a}}
=\frac{1}{\hat{a}} \frac{d\hat{a}}{d\hat{t}}\,.
\ee
The relations of the quantities 
$\hat{\bar N}, \hat{H}, \hat{\cal F}, \hat{\bar L}, \hat{L}_{,\hat{N}}$ 
with those in the JF are given in Appendix A (see also Ref.~\cite{Tsuji14}).
Using these relations as well as the background equations 
(\ref{back1})-(\ref{back2}) in the JF and the correspondence 
(\ref{backre}), we can also derive 
Eqs.~(\ref{backtra1})-(\ref{backtra2}).

The continuity equation (\ref{conti}) in the JF can be expressed in terms 
of $\hat{\rho}$ and $\hat{P}$ on account of Eq.~(\ref{backre}). 
Then, the corresponding equation in the transformed frame is given by 
\be
\frac{d\hat{\rho}}{d\hat{t}}
+3\hat{H} (\hat{\rho}+\hat{P})=Q\,,
\label{conmo}
\ee
where 
\be
Q \equiv \frac{1}{\bar{\beta}} \frac{d\bar{\beta}}{d\hat{t}} 
\hat{\rho}+3\left[ \hat{H} (1-\bar{\alpha} \bar{\beta})+
\bar{\alpha} \bar{\beta} \frac{\omega}{\hat{\bar{N}}} 
\right]\hat{P}\,.
\label{Q}
\ee
The rhs of Eq.~(\ref{conmo}) describes the coupling 
between matter and the scalar degree of freedom.

\subsubsection{Perturbation equations}
\label{pereqtra}

In the transformed frame, the perturbation equations following from 
the variations with respect to $\hat{\delta N}$, $\partial^2 \hat{\psi}$, 
and $\hat{\zeta}$ are given, respectively, by  
\ba
& &
\hspace{-0.6cm}
\left( 2\hat{\bar N}\hat{L}_{,\hat{N}}
+\hat{\bar N}^2 \hat{L}_{,\hat{N}\hat{N}}
-6\hat{H}\hat{\bar N}\hat{{\cal W}}
+12\hat{L}_{,\hat{{\cal S}}}\hat{H}^2 \right) 
\frac{\hat{\delta N}}{\hat{\bar N}} \nonumber \\
& &
\hspace{-0.6cm}
+\left( 3\frac{d\hat{\zeta}}{d\hat{t}}-\frac{\partial^2 \hat{\psi}}
{\hat{a}^2\hat{\bar N}} \right)
\hat{\bar N}\hat{{\cal W}}
-4(\hat{\bar{N}}\hat{{\cal D}}+\hat{{\cal E}})
\frac{\partial^2 \hat{\zeta}}{\hat{a}^2}=\hat{\delta \rho}\,,
\label{pereq1d} \\
& &
\hspace{-0.6cm}
\hat{{\cal W}} \frac{\hat{\delta N}}{\hat{\bar{N}}}
-\frac{4\hat{L}_{,\hat{{\cal S}}}}{\hat{\bar{N}}}
\frac{d\hat{\zeta}}{d\hat{t}}=-\hat{\delta q}\,,
\label{pereq2d} \\
& &
\hspace{-0.6cm}
\frac{1}{\hat{a}^3} \frac{d}{d\hat{t}} \left( \hat{a}^3 \hat{\bar{N}} 
\hat{{\cal Y}} \right)+4(\hat{\bar{N}}\hat{{\cal D}}
+\hat{{\cal E}}) \frac{\partial^2\hat{\delta N}}{\hat{a}^2 \hat{\bar{N}}}
+\frac{4\hat{{\cal E}}}{\hat{a}^2} \partial^2 \hat{\zeta} \nonumber \\
& &
\hspace{-0.6cm}
-3(\hat{\rho}+\hat{P}) \frac{\hat{\delta N}}{\hat{\bar{N}}}
=3\hat{\delta P}\,,
\label{pereq3d} 
\ea
where $\hat{{\cal D}}, \hat{\cal E}, \hat{{\cal W}}, \hat{{\cal Y}}$ are 
defined in the same way as Eqs.~(\ref{calD})-(\ref{W}) with  
the over-hat for background quantities. 
We can also derive Eqs.~(\ref{pereq1d})-(\ref{pereq3d}) from 
Eqs.~(\ref{pereq1})-(\ref{pereq3}) in the JF by using 
the relations (\ref{zetaE}), (\ref{Npsieq}), (\ref{appen1}) and (\ref{appen2}). 

The equation of motion for the momentum perturbation 
$\hat{\delta q}$ can be derived by employing 
Eqs.~(\ref{Npsieq}), (\ref{backre}), (\ref{deltaq}), (\ref{mure}), 
(\ref{deltaqeq}), and (\ref{appen1}), as
\be
\frac{d}{d\hat{t}} (\hat{\bar{N}} \hat{\delta q})
+3\hat{H} \hat{\bar{N}} \hat{\delta q}
=-(\hat{\rho}+\hat{P})\frac{\hat{\delta N}}{\hat{\bar{N}}}
-\hat{\delta P}\,.
\label{deltaqtra}
\ee
This is of the same form as Eq.~(\ref{deltaqeq}) in the JF.

Similarly, the density perturbation $\hat{\delta \rho}$ obeys 
\ba
\hspace{-1.0cm}
&&
\frac{d}{d\hat{t}} 
\left( \hat{\delta \rho}-\nu \hat{\rho} \hat{\delta N} \right)
+\left( 3\hat{H}-\frac{1}{\bar{\beta}}\frac{d\bar{\beta}}{d\hat{t}} \right)
\left( \hat{\delta \rho}-\nu \hat{\rho} \hat{\delta N} \right) \nonumber \\
\hspace{-1.0cm}
&&
+3\bar{\alpha} \bar{\beta} \left( \hat{H}-\frac{\omega}{\hat{\bar{N}}} 
\right) \left( \hat{\delta P}+\mu \bar{\beta} \hat{P} 
\hat{\delta N} \right) \nonumber \\
\hspace{-1.0cm}
&&
+\left( \hat{\rho}+\bar{\alpha} \bar{\beta} \hat{P} \right)
\left( 3\frac{d\hat{\zeta}}{d\hat{t}} 
-\frac{\partial^2 \hat{\psi}}{\hat{a}^2\hat{\bar N}} \right)
+\frac{\bar{\beta}\Omega^2 \hat{\bar{N}}}
{\bar{\alpha}}\frac{\partial^2 \hat{\delta q}}
{\hat{a}^2}=0\,,
\label{maEin}
\ea
the form of which is modified relative to 
Eq.~(\ref{pereq4}) in the JF.
This modification comes from an explicit coupling between 
matter and the scalar degree of freedom in the 
transformed frame.

{}From Eqs.~(\ref{pereq3d}) and (\ref{deltaqtra}) it follows that
\ba
& &
\left( \frac{d\hat{L}_{,\hat{\cal S}}}{d\hat{t}}
+\hat{H} \hat{L}_{,\hat{\cal S}}
-\frac{1}{\hat{\bar{N}}} \frac{d\hat{\bar{N}}}{d\hat{t}}
\hat{L}_{,\hat{\cal S}} \right) 
\frac{\hat{\psi}}{\hat{\bar{N}}}
+\frac{\hat{L}_{,\hat{\cal S}}}{\hat{\bar{N}}}
\frac{d\hat{\psi}}{d\hat{t}} \nonumber \\
& &+(\hat{\bar{N}}\hat{{\cal D}}+\hat{{\cal E}}) 
\frac{\hat{\delta N}}{\hat{\bar{N}}}
+\hat{{\cal E}} \hat{\zeta}=0\,.
\label{anieq2}
\ea
We introduce the gauge-invariant Bardeen potentials 
in the transformed frame, as
\be
\hat{\Psi} \equiv \frac{\hat{\delta N}}{\hat{\bar{N}}}
+\frac{d}{d\hat{t}} \left( \frac{\hat{\psi}}{\hat{\bar{N}}} 
\right)\,, \qquad
\hat{\Phi} \equiv \hat{\zeta}+\hat{H}\frac{\hat{\psi}}
{\hat{\bar{N}}}\,.
\label{PsiPhiEin}
\ee
{}From Eq.~(\ref{anieq2}) we obtain the relation similar to 
Eq.~(\ref{anire}) with additional over-hats to each quantity. 
As we see in Sec.~\ref{Einsec}, it is possible to find a metric frame 
in which some of the terms generating the anisotropic stress vanish.

We also introduce the gauge-invariant matter 
density contrast, as
\be
\hat{\delta}_m \equiv \hat{\delta}-3\hat{V}_m\,,
\label{hatdelm}
\ee
where $\hat{\delta} \equiv \delta \hat{\rho}/\hat{\rho}$ 
and $\hat{V}_m \equiv \hat{\bar{N}}\hat{H}\delta \hat{q}/\hat{\rho}$.
{}From Eqs.~(\ref{deltaqtra}) and (\ref{maEin}) we can derive the 
second-order equation for $\hat{\delta}_m$ analogous to 
Eq.~(\ref{mattereq}) in the JF. 
If we transform to the EF, the effective gravitational 
coupling with matter becomes particularly transparent.
We shall address this issue in Sec.~\ref{Einsec}.

\section{Einstein frame}
\label{Einsec}

We define the EF in which the second-order action 
of tensor perturbations $\gamma_{ij}$ is of the same 
form as that in GR \cite{Tsuji14}. 
In the following we discuss the transformation of the 
action in the JF frame to that in the EF.

\subsection{Transformation to the EF}

Expanding the action (\ref{action}) with the Lagrangian (\ref{LGLPV}) 
in terms of tensor perturbations $\gamma_{ij}$, 
the resulting quadratic action of $\gamma_{ij}$ in the JF is given by 
\be
S_2^{(h)}=\int d^4 x\,a^3 q_{\rm t} \delta^{ik} \delta^{jl}
\left( \dot{\gamma}_{ij} \dot{\gamma}_{kl}
-\frac{c_{\rm t}^2}{a^2}\partial \gamma_{ij} 
\partial \gamma_{kl} \right)\,, 
\label{L2ten}
\ee
where
\be
q_{\rm t} \equiv \frac{L_{,\cal S}}{4\bar{N}}\,,\qquad
c_{\rm t}^2 \equiv \frac{\bar{N}^2 {\cal E}}{L_{,\cal S}}\,.
\label{qt}
\ee
In Eq.~(\ref{L2ten}) the quantities $q_{\rm t}$ and $c_{\rm t}^2$ 
should be evaluated on the background, such that the kinetic 
term $X$ appearing in $L_{,\cal S}=-A_4-3HA_5$ 
and ${\cal E}=B_4+\dot{B}_5/(2\bar{N})$ corresponds to 
the time derivative $\bar{X}(t)=-\dot{\phi}^2(t)/(2\bar{N}^2)$.
We require the conditions $q_{\rm t}>0$ and 
$c_{\rm t}^2>0$ to avoid ghosts and Laplacian instabilities.
In GR we have $L=M_{\rm pl}^2R/2=
-(M_{\rm pl}^2/2) (K^2-{\cal S})+(M_{\rm pl}^2/2){\cal R}$,
in which case $q_{\rm t}=M_{\rm pl}^2/(8\bar{N})$ and 
$c_{\rm t}^2=\bar{N}^2$ 
(where $M_{\rm pl}$ is the reduced Planck mass).

Under the disformal transformation (\ref{diformaltra}) the tensor 
perturbation is invariant, see Eq.~(\ref{zetaE}). 
Hence the second-order tensor action in the 
transformed frame reads
\be
S_2^{(h)}=\int d^4 x\,\hat{a}^3 \hat{q}_{\rm t} \delta^{ik} \delta^{jl}
\left( \dot{\gamma}_{ij} \dot{\gamma}_{kl}
-\frac{\hat{c}_{\rm t}^2}{\hat{a}^2}\partial \gamma_{ij} 
\partial \gamma_{kl} \right)\,, 
\label{L2tentra}
\ee
where $\hat{q}_{\rm t} \equiv \hat{L}_{,\hat{\cal S}}/(4\hat{\bar N})$ and 
$\hat{c}_{\rm t}^2 \equiv \hat{\bar{N}}^2\hat{\cal E}
/\hat{L}_{,\hat{\cal S}}$. Comparing Eq.~(\ref{L2tentra}) 
with Eq.~(\ref{L2ten}) and using the relation (\ref{hata}), 
it follows that 
\be
\hat{q}_{\rm t}=\frac{1}{\Omega^3}q_{\rm t}\,,\qquad
\hat{c}_{\rm t}^2=\Omega^2 c_{\rm t}^2\,.
\ee

The EF corresponds to a frame in which 
both $\hat{q}_{\rm t}$ and $\hat{c}_{\rm t}^2$ are 
of the same forms as those in GR, i.e., 
$\hat{q}_{\rm t}=M_{\rm pl}^2/(8\hat{\bar N})$ and 
$\hat{c}_{\rm t}^2=\hat{\bar N}^2$. 
The tensor action (\ref{L2ten}) in the JF can be transformed 
to that in the EF for the choice
\be
\Omega^2=\frac{8q_{\rm t}c_{\rm t}}{M_{\rm pl}^2}\,,\qquad
\Gamma=\frac{8q_{\rm t}c_{\rm t}}{M_{\rm pl}^2} 
\left( \frac{c_{\rm t}^2}{\bar{N}^2}-1 \right)\frac{1}{X}\,.
\label{choice}
\ee
The quantities $q_{\rm t}$ and $c_{\rm t}^2$
in the action (\ref{L2ten}) depend on the time $t$ alone. 
Then, the factor $\Gamma$ in Eq.~(\ref{choice}) has 
the dependence $\Gamma(\phi,X)=\gamma(\phi)/X$ 
in the unitary gauge, where 
$\gamma(\phi)=(8q_{\rm t}c_{\rm t}/M_{\rm pl}^2)
(c_{\rm t}^2/\bar{N}^2-1)$.

For the choices (\ref{choice}) the terms $\alpha$ and $\beta$ in 
Eqs.~(\ref{aldef}) and (\ref{betadef}) are given by 
\be
\alpha=\frac{1}{\beta}=\Omega \frac{c_{\rm t}}{\bar{N}}\,.
\label{albere}
\ee
Since both $\alpha$ and $\beta$ are functions of $t$, 
we have $\mu=0=\nu$ from Eq.~(\ref{numu}).
Then, the coupling (\ref{Q}) reduces to 
\be
Q=-\frac{\omega}{\hat{\bar N}} \left( \hat{\rho} 
-3\hat{P} \right)-\frac{\dot{\cal C}_{\rm t}}
{\hat{\bar N}{\cal C}_{\rm t}} \hat{\rho}\,,
\label{Q2}
\ee
where 
\be
{\cal C}_{\rm t} \equiv \frac{c_{\rm t}}{\bar{N}}\,.
\label{Calct}
\ee
If $c_{\rm t}=\bar{N}$, then $\Gamma=0$ and
$Q=-(\omega/\hat{\bar{N}})( \hat{\rho}-3\hat{P})$. 
This case corresponds to the well-known conformal transformation 
arising e.g., in BD theory \cite{Amendola}. 
For radiation ($\hat{\rho}=3\hat{P}$) the coupling $Q$ 
vanishes, but for non-relativistic matter ($\hat{P}=0$), we have 
that $Q=-(\omega/\hat{\bar{N}})\hat{\rho}$.
If ${\cal C}_{\rm t}$ varies in time, the last term 
on the rhs of  Eq.~(\ref{Q2}) does not 
vanish even for radiation.

\subsection{Background equations in the EF}

The choice (\ref{choice}) corresponds to the conditions 
\ba
\hat{L}_{,\hat{\cal S}}
&=& -\hat{A}_4-3\hat{H}\hat{A}_5
=\frac{M_{\rm pl}^2}{2}\,,
\label{Lsrela}\\
\hat{\cal E}
&=&
\hat{B}_4+\frac12 \frac{d\hat{B}_5}{d\hat{t}}
=\frac{M_{\rm pl}^2}{2}\,.
\ea
Using the relation (\ref{Lsrela}), the background Eqs.~(\ref{backtra1}) 
and (\ref{backtra2}) in the EF can be written in the following forms:
\ba
3M_{\rm pl}^2 \hat{H}^2 &=& \hat{\rho}_{\rm DE}+\hat{\rho}\,,
\label{backEin1}\\
-2M_{\rm pl}^2 \frac{d\hat{H}}{d\hat{t}}
&=& \hat{\rho}_{\rm DE}+\hat{P}_{\rm DE}
+\hat{\rho}+\hat{P}\,,
\label{backEin2}
\ea
where
\ba
\hspace{-0.2cm}
\hat{\rho}_{\rm DE} 
&\equiv& 
-\hat{A}_2-6\hat{H}^3 \hat{A}_5 \nonumber \\
& &-\hat{\bar{N}} \left( \hat{A}_{2,\hat{N}}+3\hat{H}\hat{A}_{3,{\hat{N}}}
-12\hat{H}^3 \hat{A}_{5,\hat{N}} \right),\\
\hspace{-0.2cm}
\hat{P}_{\rm DE} &\equiv&
\hat{A}_2+6\hat{H}^3\hat{A}_5 \nonumber \\
& &
-\left( \frac{d\hat{A}_3}{d\hat{t}}-12\hat{H}\frac{d\hat{H}}{d\hat{t}} 
\hat{A}_5-6\hat{H}^2\frac{d\hat{A}_5}{d\hat{t}}  \right).
\ea
{}From Eqs.~(\ref{backre}) and (\ref{albere}), the matter equation of state 
$w=P/\rho$ is invariant under the transformation to the EF, i.e., 
$P/\rho=\hat{P}/\hat{\rho}$.
The energy density $\hat{\rho}_{\rm DE}$ and the pressure 
$\hat{P}_{\rm DE}$ obey the equation of motion
\be
\frac{d\hat{\rho}_{\rm DE}}{d\hat{t}}
+3\hat{H} \left( \hat{\rho}_{\rm DE}+\hat{P}_{\rm DE} 
\right)=-Q\,,
\label{rhoDEein}
\ee
where $Q$ is given by Eq.~(\ref{Q2}). 
Comparing Eq.~(\ref{conmo}) with Eq.~(\ref{rhoDEein}), 
the scalar field and matter interact with each other in the EF.

The background equations of motion in the JF
do not contain the terms $B_4$ and $B_5$.
The theories with same values of $A_{2,3,4,5}$ but with 
different $B_{4,5}$ cannot be distinguished from each other 
at the background level \cite{Kase14,Kase2}.
In other words, two theories with different values of $c_{\rm t}^2$
lead to the same background dynamics for given $A_{2,3,4,5}$.
This implies that the coupling 
$-\dot{\cal C}_{\rm t}/(\hat{\bar{N}}{\cal C}_{\rm t}) \hat{\rho}$ 
appearing in Eq.~(\ref{Q2}) does not essentially modify the 
background physics even for the theories in which 
${\cal C}_{\rm t}$ varies in time.
In Sec.~\ref{newmodelsec} we shall confirm this property for 
a concrete dark energy model. 

\subsection{Perturbations in the EF}

Substituting the relations $\hat{L}_{,\hat{\cal S}}=\hat{{\cal E}}=M_{\rm pl}^2/2$ 
into Eqs.~(\ref{pereq1d})-(\ref{anieq2}), we obtain 
the perturbation equations of motion in the EF. 
{}From Eq.~(\ref{anieq2}) the gauge-invariant Bardeen potentials 
obey the relation
\be
\hat{\Psi}+\hat{\Phi}=-\hat{\alpha}_{\rm H}\, 
\frac{\hat{\delta N}}{\hat{\bar{N}}}\,,
\label{ani2}
\ee
where $\hat{\alpha}_{\rm H}=2\hat{\bar{N}}\hat{{\cal D}}/M_{\rm pl}^2$ 
is the parameter characterizing the departure from Horndeski theories.
Since $\dot{\hat{L}}_{,\hat{\cal S}}=0$ and 
$\hat{c}_{\rm t}^2=\hat{\bar N}^2$, the first and second terms present
on the rhs of Eq.~(\ref{anire}) in the JF vanish in the EF.

The full GLPV action cannot be mapped to the full Horndeski 
action under the disformal transformation \cite{Gleyzes14,Tsuji14}, 
so the parameter $\hat{\alpha}_{\rm H}$ in Eq.~(\ref{ani2}) 
does not generally vanish. 
It is, however, possible to transform part of the GLPV action to the 
action belonging to Horndeski theories, in which case 
$\hat{\alpha}_{\rm H}=0$ and hence there is no anisotropic stress in the EF.

Let us consider perturbations of non-relativistic matter characterized 
by $P=0$ and $\delta P=0$. 
In the EF, Eqs.~(\ref{deltaqtra}) and (\ref{maEin}) reduce, respectively, to 
\ba
\hspace{-0.5cm}
& &
\frac{1}{\hat{H}} \frac{d\hat{V}_m}{d\hat{t}}
+\left( \frac{1}{\hat{H}\bar{\beta}} \frac{d\bar{\beta}}{d\hat{t}}
-\frac{1}{\hat{H}^2} \frac{d\hat{H}}{d\hat{t}} 
\right)\hat{V}_m=-\frac{\hat{\delta N}}{\hat{\bar{N}}}\,,
\label{VmeqEi} \\
\hspace{-0.5cm}
& &
\frac{d\hat{\delta}}{d\hat{t}}+3\frac{d\hat{\zeta}}{d\hat{t}}
+\frac{k^2}{\hat{a}^2} \frac{\hat{\psi}}{\hat{\bar N}}
-\frac{1}{{\cal C}_{\rm t}^2}\frac{k^2}{\hat{a}^2} 
\frac{\hat{V}_m}{\hat{H}}=0\,, 
\label{deltaEi}
\ea
where we used the background Eq.~(\ref{conmo}).
Taking the $\hat{t}$ derivative of Eq.~(\ref{deltaEi}) and 
employing Eq.~(\ref{VmeqEi}), the matter density contrast 
(\ref{hatdelm}) obeys
\ba
& &
\frac{d^2 \hat{\delta}_m}{d\hat{t}^2}+
\left( 2\hat{H}-\frac{1}{\Omega} \frac{d\Omega}{d\hat{t}}
+\frac{1}{{\cal C}_{\rm t}} \frac{d{\cal C}_{\rm t}}{d\hat{t}}
\right) \frac{d \hat{\delta}_m}{d\hat{t}}
+\frac{k^2}{\hat{a}^2} \hat{\Psi}_g \nonumber \\
& &
=-3\frac{d^2 \hat{B}}{d\hat{t}^2}
-3\left( 2\hat{H}-\frac{1}{\Omega} \frac{d\Omega}{d\hat{t}}
+\frac{1}{{\cal C}_{\rm t}} \frac{d{\cal C}_{\rm t}}{d\hat{t}}
\right)\frac{d\hat{B}}{d\hat{t}}\,,
\label{delmeqEin}
\ea
where $\hat{B} \equiv \hat{\zeta}+\hat{V}_m$, and 
\be
\hat{\Psi}_g \equiv \hat{\Psi}
-\left( \frac{1}{\Omega} \frac{d\Omega}{d\hat{t}}
-\frac{1}{{\cal C}_{\rm t}} \frac{d{\cal C}_{\rm t}}{d\hat{t}}
\right)\frac{\hat{\psi}}{\hat{\bar N}}
+\left( \frac{1}{{\cal C}_{\rm t}^2}-1 
\right) \frac{\hat{\delta N}}{\hat{\bar N}}.
\label{Psigdef}
\ee
The effective potential $\hat{\Psi}_g$ characterizes the 
strength of gravitational coupling with matter. 
In the EF, it is clear that $\hat{\Psi}_g$ is expressed in terms 
of the sum of the gravitational potential $\hat{\Psi}$ and
contributions from the variations of $\Omega$ and 
${\cal C}_{\rm t}$ as well as the difference of 
${\cal C}_{\rm t}$ from 1.  
For the theories with ${\cal C}_{\rm t}=1$, which is 
the case for BD theories, the variation of the conformal 
factor $\Omega$ gives rise to the modification to 
$\hat{\Psi}$. The deviation of ${\cal C}_{\rm t}$ from 1 and 
the variation of ${\cal C}_{\rm t}$ occur for the theories
studied in Refs.~\cite{DKT,Tsujikawa15}, 
in which case the gravitational interaction is 
modified as well.

On using the correspondence $\hat{\delta N}/\hat{\bar{N}}=\delta N/\bar{N}$ 
and $\hat{\psi}/\hat{\bar{N}}=\Omega \psi/(\bar{N}{\cal C}_{\rm t})$, 
the gravitational potential $\hat{\Psi}$ can be expressed 
by using $\Psi$ in the JF. 
Then, we obtain the simple relation 
\be
\hat{\Psi}_g=\frac{\Psi}{{\cal C}_{\rm t}^2}\,.
\label{Psigrelation}
\ee
This shows that the effective potential $\hat{\Psi}_g$ is directly 
related to $\Psi$ appearing in the matter perturbation 
Eq.~(\ref{mattereq}) in the JF. 
Once we find the evolution of $\hat{\Psi}_g$ in the EF, 
the potential $\Psi$ and the effective gravitational 
coupling $G_{\rm eff}$ in the JF are known accordingly.

\subsection{Model belonging to Horndeski theories in the EF}

One example of realizing $\hat{\alpha}_{\rm H}=0$ 
in the EF is the model described by the JF Lagrangian
\be
L=A_2(N,t)+A_3(N,t)K
+A_4(t) (K^2-{\cal S})+B_4(t){\cal R}\,.
\label{LagJFex}
\ee
In this case the tensor propagation speed squared $c_{\rm t}^2$ 
divided by $\bar{N}^2$ reads
\be
{\cal C}_{\rm t}^2=-\frac{B_4}{A_4}\,.
\ee
In Horndeski theories we have $-A_4=B_4$ from the first condition 
of Eq.~(\ref{ABcon}), but in GLPV theories there is no such restriction 
and hence ${\cal C}_{\rm t}^2$ generally differs from 1.
Since ${\cal D}=0$ and 
$\alpha_{\rm H}={\cal C}_{\rm t}^2-1$ 
for the model (\ref{LagJFex}), 
it follows that 
\be
\Psi+\Phi=
\frac{\dot{A}_4}{\bar{N}HA_4} (\zeta-\Phi)
-\left({\cal C}_{\rm t}^2-1 \right)
\left( \zeta+\frac{\delta N}{\bar{N}} \right)\,,
\ee
from Eq.~(\ref{anire}). 
For $A_4$ depending on $t$ and for ${\cal C}_{\rm t}^2$ 
different from $1$, the anisotropic stress is
present in the JF.

Under the disformal transformation with the factors
\ba
\Omega^2 &=& \frac{2\sqrt{-A_4B_4}}{M_{\rm pl}^2}\,,
\label{OmeGam0}\\
\Gamma &=& \frac{2\sqrt{-A_4B_4}}{M_{\rm pl}^2} 
\left( -\frac{B_4}{A_4}-1 \right)
\frac{1}{X}\,,
\label{OmeGam}
\ea
the Lagrangian (\ref{LagJFex}) is transformed to 
\be
\hat{L}=\hat{A}_2(\hat{N},t)+\hat{A}_3(\hat{N},t)\hat{K}
+\frac{M_{\rm pl}^2}{2} \left( \hat{{\cal S}} -\hat{K}^2+
\hat{{\cal R}} \right)\,,
\label{LagEFex}
\ee
where $\hat{A}_2=(A_2-3\omega A_3/N+6\omega^2 A_4/N^2)
/(\Omega^3 \alpha)$ and $\hat{A}_3=(A_3-4\omega A_4/N)/\Omega^3$
with $\alpha=(\sqrt{2}/M_{\rm pl}) (-B_4^3/A_4)^{1/4}$ and 
$\omega=(\dot{A}_4/A_4+\dot{B}_4/B_4)/4$.
The last term of Eq.~(\ref{LagEFex}) corresponds to the 
Einstein-Hilbert term $M_{\rm pl}^2\hat{R}/2$, 
where $\hat{R}$ is the four-dimensional Ricci scalar.
Since $\hat{\alpha}_{\rm H}=0$ for the Lagrangian (\ref{LagEFex}), 
it follows that 
\be
\hat{\Psi}+\hat{\Phi}=0\,.
\label{noani}
\ee
Thus, for the Lagrangian (\ref{LagJFex}), the disformal 
transformation allows one to de-mix the 
gravitational potentials in the EF, 
such that the anisotropy parameter 
$\hat{\eta}=-\hat{\Phi}/\hat{\Psi}$ is equivalent to 1.

While the gravitational potentials are de-mixed in the EF, the matter 
perturbation $\hat{\delta}_m$ is subject to gravitational mixing described by the 
effective potential (\ref{Psigdef}) mediated by the scalar field $\phi$.
Multiplying the term $3\hat{\bar N}\hat{H}$ for Eq.~(\ref{pereq2d}) and 
taking the sum with Eq.~(\ref{pereq1d}), one can relate $\hat{\delta}_m$ with 
metric perturbations. Let us employ the sub-horizon 
approximation \cite{subhorizon} under which the dominant contributions 
to the lhs of Eqs.~(\ref{pereq1d}) and (\ref{pereq2d}) are those involving 
$\partial^2 \hat{\psi}/\hat{a}^2$ and $\partial^2 \hat{\zeta}/\hat{a}^2$.
In the EF the terms $\hat{\cal E}$ and $\hat{\cal W}$ are given, 
respectively, by $\hat{\cal E}=M_{\rm pl}^2/2$ and 
$\hat{\cal W}=2\hat{H}M_{\rm pl}^2/\hat{\bar N}+\hat{A}_{3,\hat{N}}$. 
Provided that $2\hat{H}M_{\rm pl}^2/\hat{\bar N} \gg |\hat{A}_{3,\hat{N}}|$, 
we obtain the Poisson equation 
\be
\frac{k^2}{\hat{a}^2} \hat{\Phi} \simeq 
\frac{1}{2M_{\rm pl}^2} \hat{\rho} \hat{\delta}_m\,.
\ee
On using Eq.~(\ref{noani}) and introducing the gravitational constant 
as $G=(8\pi M_{\rm pl}^2)^{-1}$, it follows that 
\be
\frac{k^2}{\hat{a}^2} \hat{\Psi} \simeq 
-4\pi G \hat{\rho} \hat{\delta}_m\,.
\label{PsiGR}
\ee
This shows that  the gravitational coupling associated with $\hat{\Psi}$ 
is simply given by $G$ under the condition 
$2\hat{H}M_{\rm pl}^2/\hat{\bar N} \gg |\hat{A}_{3,\hat{N}}|$.\footnote{
There are some models like kinetic braidings \cite{braid} in which the dependence 
of $\hat{A}_3$ on $\hat{N}$ modifies the gravitational coupling; see 
Refs.~\cite{KimuraYa,Koba}.}
Hence, the modified gravitational interaction from GR arises from 
the terms on the rhs of Eq.~(\ref{Psigdef}) other than $\hat{\Psi}$.

In Sec.~\ref{newmodelsec} we consider a concrete model belonging 
to the Lagrangian (\ref{LagJFex}) and study the correspondence 
of physical quantities between the JF and the EF in detail.

\section{Concrete model}
\label{newmodelsec}

We study a dark energy model described by the action (\ref{action}), where 
the Lagrangian $L$ is given by Eq.~(\ref{LagJFex}). 
We consider the following functions
\ba
& &
A_2=-\frac{1}{2}\epsilon(\phi)X-V(\phi)\,,\quad
A_3=-M_{\rm pl}^2 \sqrt{-X}F_{1,\phi}\,,
\nonumber \\ & &
A_4=-\frac{1}{2}M_{\rm pl}^2 F_1(\phi)\,,\quad
B_4=\frac{1}{2}M_{\rm pl}^2 F_2(\phi)\,,
\label{AB}
\ea
where $\epsilon(\phi)$, $V(\phi)$, $F_1(\phi)$, and 
$F_2(\phi)$ are functions that depend on $\phi$, i.e., 
on $t$ in the unitary gauge. 
The dependence of $A_2$ and $A_3$ on $N$ arises in 
the kinetic term $X=-\dot{\phi}^2/N^2$.
To accommodate BD theory \cite{Brans} as well as theories 
recently proposed in Refs.~\cite{DKT,Tsujikawa15}, 
we choose the functions
\ba
& &
F_1(\phi)=e^{-2q_1 \phi/M_{\rm pl}}\,,\quad
F_2(\phi)=c_{{\rm t}i}^2e^{-2q_2 \phi/M_{\rm pl}}\,,
\nonumber \\ & &
\epsilon(\phi)=(1-6q_1^2)F_1(\phi)\,,
\ea
where $q_1, q_2, c_{{\rm t}i}$ are positive constants. 
We assume that $q_1,q_2 \ll 1$ for the compatibility 
with observations \cite{Amendola,Petto}.
In Horndeski theories, the first condition of Eq.~(\ref{ABcon}) 
demands that $F_2(\phi)$ is equivalent  to $F_1(\phi)$. 
The original BD theory without the field 
potential corresponds to the case $V(\phi)=0$ and 
$F_2(\phi)=F_1(\phi)$ with the BD parameter 
$\omega_{\rm BD}=(1-6q_1^2)/(4q_1^2)$ \cite{Uddin}.

Let us first consider theoretical consistent conditions 
in the JF. In the following we set the background value of the
lapse $\bar{N}$ to be 1. 
{}From Eqs.~(\ref{qt}) and (\ref{qs}) the conditions for avoiding 
tensor and scalar ghosts are given, respectively, by  
\ba
q_{\rm t} &=& \frac{1}{8}M_{\rm pl}^2 F_1>0\,,\\
q_{\rm s} &=& \frac{M_{\rm pl}^2 \dot{\phi}^2 F_1}
{2(M_{\rm pl}H-q_1 \dot{\phi})^2}>0\,,
\ea
which are satisfied for $F_1>0$. 
The tensor and scalar propagation speed squares 
are given, respectively, by 
\ba
c_{\rm t}^2 &=&
\frac{F_2}{F_1}=
c_{{\rm t}i}^2 e^{2(q_1-q_2)\phi/M_{\rm pl}}\,,\label{ct1}\\
c_{\rm s}^2&=&c_{\rm t}^2+\frac{(1-c_{\rm t}^2)\Omega_m}
{2x_1^2}+\frac{2(q_1-q_2)c_{\rm t}^2 (\sqrt{6}-6q_1x_1)}
{3x_1}, \label{cs1} \nonumber\\
\ea
where 
\be
x_1 \equiv \frac{\dot{\phi}}{\sqrt{6}HM_{\rm pl}}\,,\qquad
\Omega_m \equiv \frac{\rho}{3M_{\rm pl}^2H^2F_1}\,.
\ee

Under the no-ghost condition $F_1>0$, the condition (\ref{ct1}) 
is satisfied for $F_2>0$. For the theories with $q_1=q_2$, 
$c_{\rm t}^2$ is constant ($c_{\rm t}^2=c_{{\rm t}i}^2$). 
Since $c_{\rm s}^2=c_{\rm t}^2+(1-c_{\rm t}^2)\Omega_m/(2x_1^2)$ 
in this case, $c_{\rm s}^2$ is positive for $0<c_{\rm t}^2<1$, while 
$c_{\rm s}^2$ can be negative for $c_{\rm t}^2>1$. 
In Ref.~\cite{DKT} the authors studied the cosmology 
for the specific case with $q_1=q_2=0$. 
If $q_1 \neq q_2$, then
$c_{\rm t}^2$ varies in time. 
The variation of $c_{\rm t}^2$ gives rise to a
contribution to $c_{\rm s}^2$, i.e., the last term on the rhs
of Eq.~(\ref{cs1}). 
The cosmology with $q_1=0$ and $q_2 \neq 0$ was recently studied in 
Ref.~\cite{Tsujikawa15} as a model of realizing weak gravity 
on scales relevant to large-scale structures.

For dark energy models in which the ratio $\Omega_m/x_1^2$ 
decreases with time, $c_{\rm s}^2$ grows to be 
very much larger than 1 as we go back to the past.
This behavior can be avoided for 
the scaling model characterized by the potential \cite{Nunes} 
\be
V(\phi)=V_1 e^{-\lambda_1 \phi/M_{\rm pl}}+
V_2 e^{-\lambda_2 \phi/M_{\rm pl}}\,,
\label{potential}
\ee
where $V_1,V_2,\lambda_1,\lambda_2$ are positive constants. 
Provided the first potential on the rhs of 
Eq.~(\ref{potential}) dominates over the second one, 
the scaling solution with the constant ratio 
$\Omega_m/x_1^2$ is realized during radiation and 
matter eras \cite{CLW}. 
The solution exits from the scaling regime to the epoch of 
cosmic acceleration due to the existence of the second potential. 
We shall consider the situation in which 
the slopes $\lambda_1$ and $\lambda_2$ are in the range
\be
\lambda_1 \gtrsim 10\,,\qquad 
\lambda_2 \lesssim 1\,,
\ee
for consistency with the big-bang nucleosynthesis \cite{Bean} 
and the late-time cosmic acceleration \cite{CST}. 
There are seven free parameters 
$(q_1,q_2,c_{{\rm t}i},V_1,V_2,\lambda_1,\lambda_2)$ 
in our model.

\subsection{Transformation to the Einstein frame}

For the theories given by the functions (\ref{AB}), the two factors 
(\ref{OmeGam0}) and (\ref{OmeGam}) transforming 
to the EF are given, respectively, by 
\ba
\Omega^2 (\phi) &=& \sqrt{F_1(\phi)F_2(\phi)}
=F_1(\phi)c_{\rm t}(\phi)\,,\label{OmeEin} \\
\Gamma (\phi,X) &=&
\sqrt{F_1(\phi)F_2(\phi)} 
\left( \frac{F_2(\phi)}{F_1(\phi)}-1 \right) \frac{1}{X}\,.
\label{Ome}
\ea
We also have
\be
\alpha=\frac{1}{\beta}=\sqrt{F_1(\phi)c_{\rm t}^3(\phi)}\,.
\label{alEin}
\ee
Then the action in the EF is given by Eq.~(\ref{actiontra}) 
with the Lagrangian (\ref{LagEFex}), where 
\ba
\hat{A}_2 &=& \frac{\dot{\phi}^2}{2\hat{N}^2}
\left[1 -\frac32 (q_1-q_2)^2 \right]-\hat{V}(\phi)\,,\\
\hat{A}_3 &=& \frac{M_{\rm pl}\dot{\phi}}{\hat{N}} 
(q_1-q_2)\,,
\ea
and the EF frame potential
\be
\hat{V}(\phi)=\frac{V(\phi)}{\sqrt{F_1(\phi)F_2^3(\phi)}}=
\frac{V(\phi)}{c_{{\rm t}i}^3}
e^{(q_1+3q_2)\phi/M_{\rm pl}}\,.
\ee
For the JF potential (\ref{potential}) it follows that 
\be
\hat{V}(\phi)=\hat{V}_1 e^{-\mu_1 \phi/M_{\rm pl}}+
\hat{V}_2 e^{-\mu_2 \phi/M_{\rm pl}}\,,
\label{poEin}
\ee
where $\hat{V}_1=V_1/c_{{\rm t}i}^3$, 
$\hat{V}_2=V_2/c_{{\rm t}i}^3$, and 
\be
\mu_1=\lambda_1-q_1-3q_2\,,\qquad
\mu_2=\lambda_2-q_1-3q_2\,.
\ee

For the theories with $q_1=q_2$ (i.e., $c_{\rm t}^2=$\,constant) 
the term $\hat{A}_3$ vanishes, in which case
the Lagrangian (\ref{LagEFex}) corresponds to that of the 
canonical scalar field $\phi$ coupled to matter. 

\subsection{Background dynamics}

{}From Eqs.~(\ref{back1})-(\ref{conti}) the background equations 
of motion in the JF are given by 
\ba
\hspace{-0.5cm}
& & 3M_{\rm pl}^2 H^2 F_1=\frac{\epsilon}{2} \dot{\phi}^2
+V-3M_{\rm pl}^2 H\dot{\phi}F_{1,\phi}+\rho,\label{backJ1}\\
\hspace{-0.5cm}
& &-2M_{\rm pl}^2F_1 \dot{H}=\epsilon \dot{\phi}^2
+M_{\rm pl}^2 (\ddot{F}_1-H \dot{F}_1)+\rho+P,\\
\hspace{-0.5cm}
& & \dot{\rho}+3H (\rho+P)=0\,.\label{backJ3}
\ea
These equations depend on the function $F_1(\phi)$ 
but not on $F_2(\phi)$, so the quantities $q_2$ and 
$c_{{\rm t}i}^2$ are irrelevant to the background dynamics. 
This means that the theories with same $q_1$ and different 
$q_2$ (i.e., same $A_4$ and different $B_4$) cannot be 
distinguished from each other for the background 
cosmology in the JF \cite{Kase14}. 

The disformal transformation to the EF corresponds 
to the change of tensor propagation speed squared 
$c_{\rm t}^2=-B_4/A_4$ to $\hat{c}_{\rm t}^2=\hat{\bar N}^2$, 
so the quantity $q_2$ arises in the EF. 
Nevertheless we are dealing with the same physics in the 
two frames, so any physical condition (such as the stability 
of fixed points) should not be subject to change. 
In what follows we shall study the correspondence of 
background quantities between the EF and the JF.

\subsubsection{Einstein frame}

The background equations of motion in the EF 
are given by Eqs.~(\ref{backEin1})-(\ref{backEin2}), where
\ba
\hat{\rho}_{\rm DE} &=&
\frac12 \left( \frac{d\phi}{d \hat{t}} \right)^2
\left[1 -\frac32 (q_1-q_2)^2 \right]+\hat{V}(\phi) \nonumber \\
& &+3(q_1-q_2)M_{\rm pl}\hat{H} \frac{d\phi}{d \hat{t}}\,,\\
\hat{P}_{\rm DE} &=&
\frac12 \left( \frac{d\phi}{d \hat{t}} \right)^2
\left[1 -\frac32 (q_1-q_2)^2 \right]-\hat{V}(\phi) \nonumber \\
& &-(q_1-q_2)M_{\rm pl} 
\frac{d^2\phi}{d \hat{t}^2}\,.
\ea
The matter fluid and the scalar field $\phi$ obey 
Eqs.~(\ref{conmo}) and (\ref{rhoDEein}), 
respectively, with 
\be
Q=\frac{q_1+q_2}{2M_{\rm pl}}
\left( \hat{\rho}-3\hat{P} \right) \frac{d\phi}
{d\hat{t}}
-\frac{q_1-q_2}{M_{\rm pl}}\hat{\rho}
\frac{d\phi}{d\hat{t}}\,.
\label{Qein}
\ee
The first term on the rhs of Eq.~(\ref{Qein})
arises for the standard coupled dark energy 
scenario characterized by $q_1=q_2$ 
and $c_{{\rm t}i}^2=1$ \cite{Amendola}.
The second term on the rhs of Eq.~(\ref{Qein}) does 
not vanish for the theories with $q_1 \neq q_2$ 
(i.e. time-varying $c_{\rm t}^2$).

To discuss the background dynamics in the EF,
it is convenient to introduce the dimensionless quantities
\ba
\hspace{-0.8cm}
& &\hat{x}_1 \equiv \frac{1}{\sqrt{6}M_{\rm pl}\hat{H}} 
\frac{d\phi}{d\hat{t}}\,,\qquad
\hat{x}_2 \equiv \frac{\sqrt{\hat{V}}}
{\sqrt{3}M_{\rm pl}\hat{H}}\,,\nonumber \\
\hspace{-0.8cm}
& & 
\hat{\Omega}_{\rm DE} \equiv \frac{\hat{\rho}_{\rm DE}}
{3M_{\rm pl}^2 \hat{H}^2}\,,\qquad
\hat{\Omega}_m \equiv \frac{\hat{\rho}}
{3M_{\rm pl}^2 \hat{H}^2}\,, \nonumber \\
\hspace{-0.8cm}
& &w \equiv \frac{\hat{P}}{\hat{\rho}}\,,\qquad
\mu \equiv -\frac{M_{\rm pl}\hat{V}_{,\phi}}{\hat{V}}\,,
\qquad 
\hat{\epsilon}_{H} \equiv -\frac{\hat{H}'}{\hat{H}},
\ea
where a prime represents a derivative with respect to 
$\hat{{\cal N}}=\int \hat{H}d\hat{t}$.
Recall that the matter equation of state $w$ is invariant 
under the disformal transformation to the EF.
In the following we assume that $w$ is constant.
{}From Eq.~(\ref{backEin1}) there is the relation 
$\hat{\Omega}_{\rm DE}+\hat{\Omega}_m=1$.

The variables $\hat{x}_1$ and $\hat{x}_2$ obey 
the differential equations
\begin{widetext}
\ba
\hat{x}_1' &=& 
\frac{\sqrt{6}}{4} [2q_1 (3w-1)-2\sqrt{6} \{1+q_1(q_1-q_2)(3w-1)\}\hat{x}_1
+q_1\{ 3(q_1-q_2)^2-2\} (3w-1)\hat{x}_1^2 \nonumber \\
& &+2\{ 3q_2+\mu-q_1(2+3w) \}\hat{x}_2^2 ]
+\hat{x}_1\hat{\epsilon}_H\,, 
\label{dx1eq}
\\
\hat{x}_2' &=& -\frac{\sqrt{6}}{2}\mu \hat{x}_1\hat{x}_2
+\hat{x}_2\hat{\epsilon}_H\,, 
\label{dx2eq}
\\
\hat{\epsilon}_H &=& \frac32 (1+w-\hat{x}_2^2)
-\frac32 q_1 (q_1-q_2)(3w-1)+\frac32 \sqrt{6} (q_1-q_2) 
\{ 1-w+q_1(q_1-q_2)(3w-1) \}\hat{x}_1 \nonumber \\
&&-\frac34 \{ 3(q_1-q_2)^2-2 \} \{1-w+q_1(q_1-q_2)(3w-1) \}\hat{x}_1^2
-\frac32[ w-(q_1-q_2)\{ q_1(2+3w)-3q_2-\mu \}]\hat{x}_2^2\,,
\label{dotHeq}
\ea
where $\hat{\epsilon}_H$ is related to the effective equation 
of state $\hat{w}_{\rm eff}$ of the system, 
as $\hat{w}_{\rm eff}=-1+2\hat{\epsilon}_H/3$.
The field density parameter and the equation of state are given, 
respectively, by 
\ba
\hat{\Omega}_{\rm DE} &=&
1-\hat{\Omega}_m 
=\frac{\hat{x}_1}{2} \left[ 2\sqrt{6} (q_1-q_2)
+\{ 2-3(q_1-q_2)^2 \}\hat{x}_1 \right]+\hat{x}_2^2\,,\\
\hat{w}_{\rm DE}&=&
\frac{\hat{P}_{\rm DE}}{\hat{\rho}_{\rm DE}}
=\frac{3[2-3(q_1-q_2)^2]\hat{x}_1^2-6\hat{x}_2^2
-2\sqrt{6}(q_1-q_2)(\hat{x}_1'-\hat{\epsilon}_H \hat{x}_1)}
{3[2-3(q_1-q_2)^2]\hat{x}_1^2+6\hat{x}_2^2
+6\sqrt{6}(q_1-q_2)\hat{x}_1}\,.
\ea
\end{widetext}

If $\mu$ is constant, which is the case for the exponential 
potential $\hat{V}(\phi)=V_0 e^{-\mu \phi/M_{\rm pl}}$, 
there are five fixed points characterized 
by constant $\hat{x}_1$ and $\hat{x}_2$. 
They are summarized in Table~\ref{crit}. 

\begin{table*}[t]
\begin{center}
\begin{tabular}{|c|c|c|c|c|}
\hline
&  $\hat{x}_1$ & $\hat{x}_2^2$ & $\hat{w}_{\rm eff}$ &
$\hat{\Omega}_{\rm DE}$ \\
\hline
(a) & {\large $\frac{\sqrt{6}q_1(3w-1)}{3[1-w+q_1(q_1-q_2)(3w-1)]}$}
& 0 & {\large $\frac{3w(1-w)+q_1(3w-1)[q_1(6w+1)-3q_2]}
{3(1-w)+3q_1(q_1-q_2)(3w-1)}$} & Eq.\,(\ref{OmeDEa}) \\
\hline
(b1) & {\large $\frac{\sqrt{6}}{\sqrt{6}+3(q_1-q_2)}$}  & 0  & 1 & 1 \\
\hline
(b2) & {\large $-\frac{\sqrt{6}}{\sqrt{6}-3(q_1-q_2)}$}  & 0 & 1 & 1 \\
\hline
(c) & {\large $\frac{\sqrt{6}(3q_1-3q_2-\mu)}
{3[(3q_1-3q_2-\mu)(q_1-q_2)-2]}$} & 
{\large $\frac{2[6-(3q_1-3q_2-\mu)^2]}
{3[2-(q_1-q_2)(3q_1-3q_2-\mu)]^2}$} & 
{\large $-\frac{6-(3q_1-3q_2-2\mu)(3q_1-3q_2-\mu)}
{6-3(q_1-q_2)(3q_1-3q_2-\mu)}$} & 1 \\
\hline
(d) & {\large $\frac{\sqrt{6}(1+w)}{2\mu-q_1+3q_2-3w(q_1+q_2)}$ }
& {\large $\frac{2[3-3w^2-2q_1(3q_2+\mu)(3w-1)+2q_1^2(9w^2+3w-2)]}
{[2\mu-q_1+3q_2-3w(q_1+q_2)]^2}$}  & 
{\large $\frac{q_1+3q_2(w-1)+(3q_1+2\mu)w}
{2\mu-q_1+3q_2-3(q_1+q_2)w}$} & Eq.\,(\ref{OmeDEd}) \\
\hline
\end{tabular}
\end{center}
\caption[crit]{The fixed points in the EF and corresponding 
values of $\hat{w}_{\rm eff}$ and $\hat{\Omega}_{\rm DE}$
for the system characterized by the autonomous 
Eqs.~(\ref{dx1eq})-(\ref{dx2eq}) with Eq.~(\ref{dotHeq}).
The scaling radiation and matter points (d) are stable for  
$\mu \gtrsim 10$ and $q_1,q_2 \ll 1$, whereas the accelerated 
point (c) is stable for $\mu \lesssim 1$ and $q_1,q_2 \ll 1$. 
The potential (\ref{poEin}) allows for the transition from 
the matter point (d) with $\mu \simeq \mu_1 \gtrsim 10$ 
to the point (c) with $\mu \simeq \mu_2 \lesssim 1$. 
}
\label{crit} 
\end{table*}

For the fixed point (a), the field density parameter reads
\begin{widetext}
\be
\hat{\Omega}_{\rm DE}=\frac
{q_1(3w-1)[q_1\{ 4-3(q_1-q_2)^2\}-6q_2+
3\{3q_1(q_1-q_2)^2+2q_2\}w]}{3[1-w+q_1(q_1-q_2)(3w-1)]^2}\,, 
\label{OmeDEa}
\ee
\end{widetext}
so that both $\hat{\Omega}_{\rm DE}$ and $\hat{x}_1$ 
vanish for radiation ($w=1/3$).
If $q_2=q_1$, it follows that 
$\hat{x}_1=\sqrt{6}q_1(3w-1)/[3(1-w)]$, 
$\hat{w}_{\rm eff}=[2q_1^2(1-3w)^2+3w(1-w)]/[3(1-w)]$, and 
$\hat{\Omega}_{\rm DE}=2q_1^2(3w-1)^2/[3(w-1)^2]$ for 
the point (a).
When $w=0$, this corresponds to the $\phi$-matter-dominated 
era ($\phi$MDE) \cite{Amendola} characterized by 
$\hat{w}_{\rm eff}=\hat{\Omega}_{\rm DE}=2q_1^2/3$.
Provided that $w \neq 1/3$, the point (a) is a kind of scaling 
solution with a constant ratio 
$\hat{\Omega}_m/\hat{\Omega}_{\rm DE}$.

Since the effective equations of state $\hat{w}_{\rm eff}$ for
the points (b1) and (b2) are 1, they are neither relevant to 
radiation/matter eras nor the late-time cosmic acceleration. 

The point (c) is the scalar-field dominated 
point ($\hat{\Omega}_{\rm DE}=1$) 
relevant to dark energy. When $q_2=q_1$ 
we have $\hat{x}_1=\mu/\sqrt{6}$, 
$\hat{x}_2=\sqrt{1-\mu^2/6}$, and 
$\hat{w}_{\rm eff}=-1+\mu^2/3$, 
so the cosmic acceleration occurs for $\mu^2<2$. 
For $q_2 \neq q_1$, $\hat{w}_{\rm eff}$ is close to $-1$
provided that $q_1, q_2, \mu$ are smaller than 
the order of 1.

The point (d) corresponds to the scaling solution with 
the field density parameter
\begin{widetext}
\be
\hat{\Omega}_{\rm DE}=
\frac{q_1^2(9w^2-30w-23)+2q_1[8\mu+3q_2 \{9+w(4+3w) \}]
+3(w+1)[4-4q_2\mu+3q_2^2(w-3)]}
{[2\mu-q_1+3q_2-3w(q_1+q_2)]^2}\,.
\label{OmeDEd}
\ee
\end{widetext}
When $q_2=q_1$, we have that 
$\hat{\Omega}_{\rm DE}=
[3(1+w)-q_1(3w-1)(\mu+q_1-3q_1w)]/(\mu+q_1-3q_1w)^2$ 
and $\hat{w}_{\rm eff}=[\mu w+q_1(3w-1)]/[\mu-q_1(3w-1)]$.
In this case the radiation scaling solution corresponds to 
$\hat{\Omega}_{\rm DE}=4/\mu^2$ and $\hat{w}_{\rm eff}=1/3$, 
whereas the matter scaling solution is characterized by 
$\hat{\Omega}_{\rm DE}=(3+q_1 \mu+q_1^2)/(\mu+q_1)^2$ 
and $\hat{w}_{\rm eff}=-q_1/(\mu+q_1)$.
If the fixed point (d) is stable, the solutions approach it during the 
radiation and matter eras.

The stability of the above fixed points can be analyzed by considering 
small perturbations $\delta \hat{x}_1$ and $\delta \hat{x}_2$ 
about each of them. The linearized version of 
Eqs.~(\ref{dx1eq}) and (\ref{dx2eq}) 
can be written in the form
\ba
\left(
\begin{array}{c}
\delta \hat{x}_1' \\
\delta \hat{x}_2'
\end{array}
\right) = {\cal M} \left(
\begin{array}{c}
\delta \hat{x}_1 \\
\delta \hat{x}_2
\end{array}
\right) \,,
\ea
where ${\cal M}$ is a $2 \times 2$ matrix. 
If the two eigenvalues $\hat{\kappa}_{1,2}$ of ${\cal M}$ are 
negative (or imaginary with negative real parts), 
then the corresponding fixed point is stable.

In the presence of radiation ($w=1/3$), the eigenvalues 
of the point (d) are given by 
\be
\hat{\kappa}_{1,2}^{(\rm d)}=-\frac{\mu-3q_1+3q_2 
\pm \sqrt{64-15(\mu-3q_1+3q_2)^2}}
{2(\mu-q_1+q_2)}\,.
\label{eigen1}
\ee
For $\mu \gtrsim 10$ and $q_1,  q_2 \ll 1$, $\hat{\kappa}_{1,2}^{({\rm d})}$ 
are imaginary with negative real parts, so the point (d) is a stable spiral. 
For non-relativistic matter ($w=0$),  
the eigenvalues of the point (d) read
\be
\hat{\kappa}_{1,2}^{(\rm d)}=
-\frac{3(\mu-q_1+3q_2) \pm \sqrt{D_{(\rm d)}}}
{2(2\mu-q_1+3q_2)}\,,
\label{eigen2}
\ee
where 
$D_{(\rm d)}=9(\mu-q_1+3q_2)^2-24[3-4q_1^2+2q_1(\mu+3q_2)]
[(\mu+3q_2)^2-5q_1(\mu+3q_2)-3+6q_1^2]$.
For $\mu \gtrsim 10$ and $q_1, q_2 \ll 1$, the eigenvalues (\ref{eigen2}) 
are again imaginary with negative real parts.
Thus, the first potential on the rhs of Eq.~(\ref{poEin}) leads to 
the scaling radiation and matter eras driven by the fixed point (d) 
with $\mu=\mu_1$.

The point (a) can be potentially relevant to radiation and matter 
eras, but one of the eigenvalues is positive, i.e., 
$\hat{\kappa}_{1}^{(\rm a)}=2$ for $w=1/3$ and 
$\hat{\kappa}_{1}^{(\rm a)}=(3+2q_1\mu+6q_1q_2-4q_1^2)
/[2(1+q_1q_2-q_1^2)]$ 
for $w=0$. Hence the solutions are attracted by the scaling solution (d) 
rather than the point (a). For $\mu$ smaller than the order of 1, 
the stable scaling matter solution (d) with $\Omega_{\rm DE}<1$ does not 
exist \cite{Tsuji2006}, in which case the matter era is replaced 
by the $\phi$MDE (a) \cite{Amendola}.
In our model,  we do not consider this latter case to avoid very large values of
$c_{\rm s}^2$ in the early cosmological epoch.

After the dominance of the second potential on the rhs of Eq.~(\ref{poEin}),  
the solutions exit from the scaling matter era (d) to the epoch of 
cosmic acceleration driven by the point (c).
In the presence of non-relativistic matter, the eigenvalues of 
the point (c) are given by 
\ba
\hspace{-0.4cm}
\hat{\kappa}_{1}^{(\rm c)}
&=& -\frac{6-(\mu-3q_1+3q_2)^2}{2+(q_1-q_2)(\mu-3q_1+3q_2)}\,,
\label{eigen3} \\
\hspace{-0.4cm}
\hat{\kappa}_{2}^{(\rm c)}
&=&
-\frac{6-2(\mu-3q_1+3q_2)(\mu-2q_1+3q_2)}
{2+(q_1-q_2)(\mu-3q_1+3q_2)}\,.
\label{eigen4}
\ea
For $\mu \lesssim 1$ and $q_1,q_2 \ll 1$,  it is clear that 
both $\hat{\kappa}_{1}^{(\rm c)}$ and 
$\hat{\kappa}_{2}^{(\rm c)}$ are negative. 
Hence, the solutions finally approach the accelerated 
attractor (c) with $\hat{w}_{\rm eff}$
close to $-1$ for $\mu_2 \ll 1$ 
(see Table \ref{crit} for the value of $\hat{w}_{\rm eff}$).

In summary, for the potential (\ref{poEin}) with $\mu_1 \gtrsim 10$ and 
$\mu_2 \lesssim 1$, the background cosmological sequence in the 
EF is as follows:
(i) scaling radiation point (d) with $w=1/3$ and $\mu=\mu_1$, $\to$
(ii) scaling matter point (d) with $w=0$ and $\mu=\mu_1$, $\to$
(iii) accelerated point (c) with $\mu=\mu_2$.

\subsubsection{Jordan frame}

The background dynamics in the JF can be known by 
using relations for physical quantities between the two frames.
We define the dimensionless quantities
\ba
& &
x_1 \equiv \frac{\dot{\phi}}{\sqrt{6}M_{\rm pl}H}\,,\qquad
x_2 \equiv \frac{\sqrt{V}}{\sqrt{3F_1}M_{\rm pl}H}\,,
\nonumber \\
& &
\lambda \equiv -\frac{M_{\rm pl}V_{,\phi}}{V}\,,\qquad
\Omega_m \equiv \frac{\rho}{3M_{\rm pl}^2H^2F_1}\,,
\label{lamdef}
\ea
where the field density parameter is given by 
$\Omega_{\rm DE}=1-\Omega_{m}$.
On using Eqs.~(\ref{backre}), (\ref{OmeEin}), (\ref{alEin}), 
and (\ref{appen1}), we obtain the following correspondence
\ba
& &
x_1=(1+\omega_H)\hat{x}_1\,,\qquad
x_2=(1+\omega_H)\hat{x}_2\,,\nonumber \\
& &
\Omega_m=(1+\omega_H)^2\hat{\Omega}_m\,,
\label{backcore}
\ea
where 
\be
\omega_H \equiv \frac{\dot{\Omega}}{H\Omega}
=-\frac{\sqrt{6}}{2} (q_1+q_2)x_1
=-\frac{\sqrt{6} (q_1+q_2) \hat{x}_1}{2+\sqrt{6} (q_1+q_2) \hat{x}_1}\,.
\label{omegaH}
\ee
The slow-roll parameter $\epsilon_H \equiv -\dot{H}/H^2$, which 
is associated with the effective equation of state $w_{\rm eff}$ in the 
JF as $w_{\rm eff}=-1+2\epsilon_H/3$, satisfies the relation
\be
\epsilon_H=(1+\omega_H) \left[ \hat{\epsilon}_H
-\frac{\sqrt{6}}{2} (q_1-3q_2)\hat{x}_1
+\frac{1}{\hat{H}(1+\omega_H)} 
\frac{d\omega_{H}}{d\hat{t}} \right]\,.
\label{epcore}
\ee
The slope $\lambda$ defined in Eq.~(\ref{lamdef}) is related to 
the slope $\mu$ in the EF, as
\be
\mu=\lambda-q_1-3q_2\,.
\label{mulam}
\ee

Using the above correspondence, one can readily translate 
the background cosmological dynamics in the EF to that in the JF. 
In the JF, the fixed point (d) in Table \ref{crit} corresponds to 
\ba
\hspace{-0.9cm}
& &
x_1^{\rm (d)} = \frac{\sqrt{6}(1+w)}{2\lambda}\,,\nonumber \\
\hspace{-0.9cm}
& &
x_2^{\rm (d)} = \frac{\sqrt{3(1-w^2)+2q_1(1-3w)[\lambda-3q_1(1+w)]}}
{\sqrt{2}\lambda},
\ea
with 
\ba
w_{\rm eff}^{\rm (d)} &=& w-2(1+w) \frac{q_1}{\lambda}\,,\\
\Omega_{\rm DE}^{\rm (d)} &=& 
\frac{3(1+w)(1-4q_1^2)+q_1\lambda (7+3w)}
{\lambda^2}\,.
\ea
The fixed point (c) translates to 
\ba
x_1^{\rm (c)} &=& \frac{\lambda-4q_1}
{\sqrt{6}(1-4q_1^2+q_1\lambda)}\,,\nonumber \\
x_2^{\rm (c)} &=& \frac{\sqrt{1-(\lambda-4q_1)^2/6}}
{1-4q_1^2+q_1\lambda},
\ea
with 
\ba
w_{\rm eff}^{\rm (c)} &=& 
-\frac{3-\lambda^2-20q_1^2+9q_1\lambda}
{3(1-4q_1^2+q_1\lambda)}\,,\\
\Omega_{\rm DE}^{\rm (c)} &=& 1\,.
\ea
The quantity $q_2$ disappears after the transformation from 
the EF to the JF, which reflects the fact that the factor $F_2$ 
is absent in the background equations (\ref{backJ1})-(\ref{backJ3}).

The stability of fixed points should be independent of the values of 
$q_2$. Substituting Eq.~(\ref{mulam}) into Eqs.~(\ref{eigen1})-(\ref{eigen4}), 
it follows that the numerators of the eigenvalues do not contain the term $q_2$. 
In the denominators there are still $q_2$-dependent terms, 
but they are simply associated with the transformation of
the number of $e$-foldings, i.e., 
\be
\frac{d\hat{\cal N}}{d{\cal N}}=
1-\frac{\sqrt{6}}{2}(q_1+q_2)x_1\,.
\label{dcalN}
\ee
The evolution of homogenous perturbations 
$\delta \hat{x}_j \propto e^{\hat{\kappa}_j \hat{{\cal N}}}$ ($j=1,2$) 
in the EF translates to the JF evolution 
proportional to $e^{\kappa_j {\cal N}}$, where 
$\kappa_j=\hat{\kappa}_j [1-\sqrt{6}(q_1+q_2)x_1/2]$. 
On using the index $\kappa_j$, the $q_2$ dependence in the 
denominators of $\hat{\kappa}_j$ vanishes identically.
Provided the rhs of Eq.~(\ref{dcalN}) is positive, which 
is the case for $q_1,q_2 \ll 1$ and $|x_1| \lesssim 1$, the stability 
conditions of fixed points are identical to each other 
in the two frames.

The above discussion shows that, in the JF, the scaling 
radiation fixed point (d) with $w=1/3$ and $\lambda=\lambda_1$  
is followed by the scaling matter point (d) with $w=0$ and $\lambda=\lambda_1$, 
and then the solutions finally approach the point (c) 
with $\lambda=\lambda_2$. 
During this sequence, the effective equation of 
state and the field density parameter evolve as 
(i) $w_{\rm eff}=1/3-8q_1/(3\lambda_1)$, 
$\Omega_{\rm DE}=4(1-4q_1^2+2q_1\lambda_1)/\lambda_1^2$ (radiation era), 
$\to$
(ii) $w_{\rm eff}=-2q_1/\lambda_1$, 
$\Omega_{\rm DE}=[3(1-4q_1^2)+7q_1\lambda_1]/\lambda_1^2$ (matter era),
$\to$
(iii)  $w_{\rm eff}=-(3-\lambda_2^2-20q_1^2+9q_1\lambda_2)/
[3(1-4q_1^2+q_1\lambda_2)]$, 
$\Omega_{\rm DE}=1$ (accelerated era).

\subsection{Perturbations and matter-scalar couplings}

We consider the evolution of cosmological perturbations and the resulting 
matter-scalar coupling in the presence of non-relativistic matter satisfying 
$P=0$ and $\delta P=0$.
In what follows we shall focus on the case where
\be
q_1=q_2 \equiv q\,,
\label{qeq}
\ee
under which $c_{\rm t}^2$ is constant. 
Then, at the background level, the coupling (\ref{Qein}) reduces to 
\be
Q=\frac{q\hat{\rho}}{M_{\rm pl}} 
\frac{d\phi}{d\hat{t}}\,.
\ee
The quantity $\omega_H$ in Eq.~(\ref{omegaH}) reads
\be
\omega_H=\frac{\dot{F}_1}{2HF_1}=\frac{\dot{\alpha}}{H\alpha}
=-\frac{\dot{\beta}}{H\beta}
=-\frac{q\dot{\phi}}{HM_{\rm pl}}\,.
\ee

In the EF, the gauge-invariant matter density contrast is 
defined by Eq.~(\ref{hatdelm}).
Since $\hat{\delta}=\delta$ and $\hat{V}_m=(1+\omega_H)V_m$, 
$\hat{\delta}_m$ is not equivalent to $\delta_m$. 
For the perturbation deep inside the Hubble radius the velocity potential 
$\hat{V}_m$ is much smaller than $\hat{\delta}$, so the difference between 
$\hat{\delta}_m$ and $\delta_m$ is small.
We introduce the effective 
gravitational coupling $\hat{G}_{\rm eff}$ in the EF, as
\be
\frac{k^2}{\hat{a}^2} \hat{\Psi}_g=
-4\pi \hat{G}_{\rm eff} \hat{\rho} \hat{\delta}_m\,,
\label{PoEin}
\ee
where $\hat{\Psi}_g$ is given by Eq.~(\ref{Psigdef}).
For the choice (\ref{qeq}), $\hat{\Psi}_g$ reduces to 
\be
\hat{\Psi}_g=\hat{\Psi}+\frac{q\phi'}{M_{\rm pl}}\hat{\chi}
+\left( \frac{1}{c_{\rm t}^2}-1 \right) 
\frac{\delta \hat{N}}{\hat{\bar N}}\,,
\label{Psigex}
\ee
where
\be
\hat{\chi} \equiv \frac{\hat{H} \hat{\psi}}
{\hat{\bar N}}\,.
\ee
Using the relation (\ref{Psigrelation}) as well as the approximation 
$\hat{\delta}_m \simeq \delta_m$ for the perturbations deep inside 
the Hubble radius, we can rewrite Eq.~(\ref{PoEin}) 
of the form $(k^2/a^2)\Psi \simeq -4\pi (\hat{G}_{\rm eff}/F_1)\rho \delta_m$. 
Hence, the effective gravitational coupling $G_{\rm eff}$ in the JF is related to 
$\hat{G}_{\rm eff}$, as 
\be
G_{\rm eff} \simeq \frac{\hat{G}_{\rm eff}}{F_1}\,.
\ee
The above discussion shows that, once $\hat{\Psi}_g$ is known 
by solving the perturbation equations in the EF, the gravitational 
potential $\Psi$ and the resulting matter-scalar coupling $G_{\rm eff}$ 
in the JF are determined accordingly.

Since $\hat{\alpha}_{\rm H}=0$ in the EF, 
we have the relation (\ref{noani}), i.e., 
\be
\hat{\Phi}=-\hat{\Psi}\,.
\ee
On using the relations $\hat{\delta N}/\hat{\bar{N}}=\delta N/\bar{N}$, 
$\hat{\zeta}=\zeta$, and $\chi \equiv H \psi/\bar{N}=c_{\rm t}^2 \hat{\chi}/(1+\omega_{H})$ 
with $\bar{N}=1$, the gravitational potentials $\Psi$ and $\Phi$ 
in the JF are related to $\hat{\Psi}$ and $\hat{\Phi}$ in the EF, as 
\ba
\Psi &=&
\hat{\Psi}+\left( c_{\rm t}^2 -1 \right) \frac{d}{d\hat{t}} 
\left( \frac{\hat{\chi}}{\hat{H}} \right)-c_{\rm t}^2 
\frac{\omega_{H}}{1+\omega_{H}} \hat{\chi}\,,
\label{Psitra}\\
\Phi &=&
\hat{\Phi}+\left( \frac{c_{\rm t}^2}{1+\omega_{H}}
-1 \right) \hat{\chi}\,.
\label{Phitra}
\ea
The anisotropy parameter $\eta$ in the JF generally differs from 1 
due to the presence of the perturbation $\hat{\chi}$. 

\subsubsection{Einstein frame}

Let us study the evolution of perturbations in the EF 
during the scaling matter and accelerated epochs.
{}From Eqs.~(\ref{pereq1d})-(\ref{anieq2}) 
we obtain the perturbation equations 
\ba
&&\hat{\zeta}' = \frac{\hat{\delta N}}{\hat{\bar N}}+\frac32 
\hat{\Omega}_m \hat{V}_m\,, 
\label{zeeq} \\
&&\hat{\chi}' = -(\hat{\epsilon}_H+1)\hat{\chi} 
-\hat{\zeta}-\frac{\hat{\delta N}}{\hat{\bar N}}\,,
\label{chieq} \\
&&\hat{\delta}'=-3\hat{\zeta}'-\hat{\cal K}^2
\left( \hat{\chi}-\frac{\hat{V}_m}{c_{\rm t}^2} 
\right)\,,
\label{conper2} \\
&&\hat{V}_m'=-\left( \hat{\epsilon}_H
+\sqrt{6}q\hat{x}_1 \right) \hat{V}_m
-\frac{\delta \hat{N}}{\hat{\bar N}}\,,
\label{conper1}\\
& &\frac{\hat{\delta N}}{\hat{\bar N}}
= \frac{3\hat{\Omega}_m \hat{\delta}_m
-2\hat{\cal K}^2 (\hat{\chi}+\hat{\zeta})}
{6 \hat{x}_1^2}\,,
\ea
where $\hat{\cal K} \equiv k/(\hat{a}\hat{H})$.
Taking the $\hat{{\cal N}}$ derivative of Eq.~(\ref{conper1}) and 
using other equations of motion, the velocity 
potential $\hat{V}_m$ obeys
\ba
&&
\hat{V}_m''+{\cal C}_1 \hat{V}_m'+\left[ 
\frac{(1-c_{\rm t}^2) \hat{\Omega}_m}{2c_{\rm t}^2 \hat{x}_1^2} 
\hat{\cal K}^2+{\cal C}_2 \right] \hat{V}_m \nonumber \\
&&=-\left[ \hat{\chi}+\frac{\sqrt{6}q}{3\hat{x}_1}
\left( \hat{\chi}+\hat{\zeta} \right) \right]\hat{\cal K}^2\,,
\label{Vmseeq}
\ea
where 
\ba
{\cal C}_1
&=& [2(3\hat{x}_1-\sqrt{6}\mu)(\hat{x}_1^2-1) \nonumber \\
& &+\hat{\Omega}_m \{3\hat{x}_1-2\sqrt{6}(\mu+q)\}]
/(2\hat{x}_1)\,,\\
{\cal C}_2
&=&3[12\hat{x}_1^5-4\sqrt{6}\mu \hat{x}_1^4
-2\hat{x}_1^3(9+2q^2+3q\mu-6\hat{\Omega}_m)  \nonumber \\
& &+3\hat{x}_1 \{2q\mu+\hat{\Omega}_m 
(\hat{\Omega}_m-1-2q(q+\mu)) \} \nonumber \\
& &
-\sqrt{6}\hat{x}_1^2 \{ 4q-4\mu+(3q+
5\mu)\hat{\Omega}_m \} \nonumber \\
& &-\sqrt{6}(q+\mu)\hat{\Omega}_m(\hat{\Omega}_m-1) 
]/(2\hat{x}_1)\,.
\ea

The general solution to Eq.~(\ref{Vmseeq}) can be written 
in the form $\hat{V}_m=\hat{V}_m^{(s)}+\hat{V}_m^{(h)}$, where 
$\hat{V}_m^{(s)}$ is the special solution and $\hat{V}_m^{(h)}$ 
is the homogenous solution derived by setting the rhs
of Eq.~(\ref{Vmseeq}) to be zero. 
For the sub-horizon perturbations satisfying $\hat{\cal K} \gg 1$, 
the homogenous solution induces the rapid oscillation of 
the velocity potential with a frequency associated with the term 
$(1-c_{\rm t}^2) \hat{\Omega}_m \hat{\cal K}^2/(2c_{\rm t}^2\hat{x}_1^2)$.

For the theories with $q=0$, as long as the homogenous solution is 
initially suppressed relative to the special solution, 
it was found in Ref.~\cite{DKT} that the perturbations 
$\chi$, $\zeta$, and $V_m$ in the JF stay nearly constant 
during the scaling matter epoch.
As we confirm later numerically, this is also the case for $q \neq 0$.
At the scaling fixed point (d) with 
$w=0$ and $\mu=\mu_1$, the ratio $\hat{\Omega}_m/\hat{x}_1^2$ 
is constant. We consider the sub-horizon perturbations 
satisfying the condition 
$(1-c_{\rm t}^2) \hat{\Omega}_m\hat{\cal K}^2/(2c_{\rm t}^2\hat{x}_1^2) 
\gg |{\cal C}_2|$.
Then, the special solution to Eq.~(\ref{Vmseeq}) is given by 
\be
\hat{V}_m^{(s)}
\simeq -\left[ \hat{\chi}+\frac{\sqrt{6}q}{3\hat{x}_1}
\left( \hat{\chi}+\hat{\zeta} \right) \right]
\frac{2c_{\rm t}^2 \hat{x}_1^2}{(1-c_{\rm t}^2)
\hat{\Omega}_m}\,.
\label{Vmspe}
\ee
Let us derive solutions along which $\hat{\zeta}$ and $\hat{\chi}$ stay 
nearly constant in the scaling matter era. 
Setting $\hat{\zeta}' \simeq 0$ and $\hat{\chi}' \simeq 0$ in 
Eqs.~(\ref{zeeq}) and (\ref{chieq}), respectively, we obtain 
\ba
& &\hspace{-0.8cm} \hat{\chi}
\simeq -\frac{3[1-c_{\rm t}^2 (1-\sqrt{6}q \hat{x}_1)]\hat{\Omega}_m}
{2\hat{\cal K}^2[3c_{\rm t}^2 \hat{x}_1^2
+(1-c_{\rm t}^2)\hat{\epsilon}_H]} 
\hat{\delta}_m,\label{perana1}\\
& &\hspace{-0.8cm} \hat{\zeta}
\simeq \frac{3 
[(1+\hat{\epsilon}_H)(1-c_{\rm t}^2)+c_{\rm t}^2 \hat{x}_1 
(3\hat{x}_1+\sqrt{6}q)]\hat{\Omega}_m}
{2\hat{\cal K}^2[3c_{\rm t}^2 \hat{x}_1^2
+(1-c_{\rm t}^2)\hat{\epsilon}_H]} \hat{\delta}_m,\\
& &\hspace{-0.8cm}\hat{V}_m^{(s)} 
\simeq \frac{c_{\rm t}^2 \hat{x}_1 (3\hat{x}_1
-\sqrt{6} q \hat{\epsilon}_H)}
{\hat{\cal K}^2[3c_{\rm t}^2 \hat{x}_1^2
+(1-c_{\rm t}^2)\hat{\epsilon}_H]}\hat{\delta}_m,
\label{perana3}\\
& &\hspace{-0.8cm}
\frac{\hat{\delta N}}{\hat{\bar N}}
\simeq -\frac{3c_{\rm t}^2\hat{x}_1 (3\hat{x}_1-\sqrt{6}q
\hat{\epsilon}_H)\hat{\Omega}_m}
{2\hat{\cal K}^2[3c_{\rm t}^2 \hat{x}_1^2
+(1-c_{\rm t}^2)\hat{\epsilon}_H]} \hat{\delta}_m,
\label{perana4}
\ea
where, in the denominators of Eqs.~(\ref{perana1})-(\ref{perana4}), 
we have neglected the terms without containing $\hat{\cal K}^2$. 
The gravitational potentials $\hat{\Psi}$ and 
$\hat{\Phi}$, which are defined by Eq.~(\ref{PsiPhiEin}), obey
\be
\hat{\Phi}=-\hat{\Psi} \simeq 
\frac{3\hat{\Omega}_m}{2\hat{\cal K}^2} \hat{\delta}_m\,,
\label{PhiGR}
\ee
which is equivalent to Eq.~(\ref{PsiGR}) with Eq.~(\ref{noani}).
Since $\hat{A}_3=0$ for $q_1=q_2$, the condition 
$2\hat{H}M_{\rm pl}^2/\hat{\bar N} \gg |\hat{A}_{3,{\hat{N}}}|$ 
used for the derivation of Eq.~(\ref{PsiGR}) is automatically satisfied.

Substituting Eqs.~(\ref{perana1}), (\ref{perana4}), and (\ref{PhiGR}) 
into Eq.~(\ref{Psigex}), we obtain
\ba
\hspace{-0.5cm}
\hat{\Psi}_g 
&\simeq& -\frac{3}{2\hat{\cal K}^2} \hat{\Omega}_m \hat{\delta}_m
\biggl[ 1+2q^2 \nonumber \\
\hspace{-0.5cm}
& &+ \frac{(1-c_{\rm t}^2)(\sqrt{6}\hat{x}_1+2q)
(\sqrt{6}\hat{x}_1-2q\hat{\epsilon}_H)}
{2\{3c_{\rm t}^2 \hat{x}_1^2
+(1-c_{\rm t}^2)\hat{\epsilon}_H\}} \biggr],
\label{PsigEF}
\ea
where $\hat{x}_1=\sqrt{6}/[2(\mu+q)]$ and 
$\hat{\epsilon}_H=3\mu/[2(\mu+q)]$ 
for the fixed point (d) with $w=0$.
The effective gravitational coupling 
$\hat{G}_{\rm eff}$ during the scaling matter 
epoch reads 
\be
\frac{\hat{G}_{\rm eff}}{G} 
\simeq 1+2q^2+
\frac{(1-c_{\rm t}^2)(\sqrt{6}\hat{x}_1+2q)
(\sqrt{6}\hat{x}_1-2q\hat{\epsilon}_H)}
{2\{3c_{\rm t}^2 \hat{x}_1^2
+(1-c_{\rm t}^2)\hat{\epsilon}_H\}},
\label{Geffex}
\ee
where the term $2q^2$ arises in BD theories after the 
conformal transformation to the EF \cite{Amendola}.
The last term on the rhs of Eq.~(\ref{Geffex}) does not 
vanish for $c_{\rm t}^2$ different from 1. 
Since we are now considering the case where $c_{\rm t}^2$ 
is constant, the variation of $c_{\rm t}^2$ does not 
appear in the expression of Eq.~(\ref{Geffex}).

For the perturbations deep inside the Hubble radius, 
the rhs of 
Eq.~(\ref{delmeqEin}) can be neglected relative to 
the lhs of it. Then, during the scaling matter era, the 
matter perturbation obeys
\be
\hat{\delta}_m''+\frac{\mu-2q}{2(\mu+q)}\hat{\delta}_m'
-\frac32 \frac{\hat{G}_{\rm eff}}{G} 
\hat{\Omega}_m \hat{\delta}_m \simeq 0\,.
\label{delmso}
\ee
Provided that $\mu \gg q$, there is a growing-mode 
solution to Eq.~(\ref{delmso}),
\be
\hat{\delta}_m \propto 
\hat{a}^p,\qquad
p=\frac{\mu-2q}{4(\mu+q)} \left[ 
\sqrt{1+\frac{24(\mu+q)^2}{(\mu-2q)^2}\hat{g}}-1 \right],
\ee
and $\hat{g} \equiv (\hat{G}_{\rm eff}/G)\hat{\Omega}_m$.
For $q/\mu \to 0$ and $\hat{g} \to 1$, the matter density contrast 
evolves as $\hat{\delta}_m \propto \hat{a}$. 
In this case, the quantity $(\hat{a}\hat{H})^2\hat{\delta}_m$, 
which appears in Eqs.~(\ref{perana1})-(\ref{perana4}), 
remains constant, so the perturbations $\hat{\chi}$, $\hat{\zeta}$, 
$\hat{V}_m^{(s)}$, and $\hat{\delta N}/\hat{\bar N}$
do not vary in time.

For $q \neq 0$ and $c_{\rm t}^2 \neq 1$, the quantity $\hat{g}$ 
is different from 1. 
As long as $q \ll 1$ and $\mu \gg q$, the deviation of 
$\hat{g}$ from 1 is small, so the analytic 
formulas (\ref{perana1})-(\ref{perana4}) are approximately valid in the 
scaling matter era. The perturbations $\hat{\chi}$ and 
$\hat{\zeta}$ start to vary after the Universe enters the epoch of 
cosmic acceleration, in which regime the analytic solutions 
(\ref{perana1})-(\ref{perana4}) are no longer valid.

\begin{figure}
\includegraphics[height=3.2in,width=3.3in]{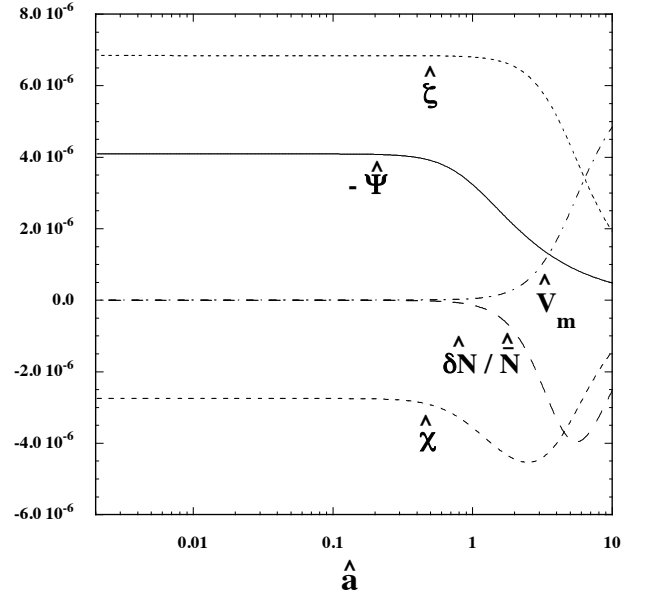}
\caption{\label{zetafig}
Evolution of the perturbations $\hat{\zeta}$, $\hat{\chi}$, 
$\hat{V}_m$, $\hat{\delta N}/\hat{\bar N}$, 
and $-\hat{\Psi}$ in the EF versus the scale factor $\hat{a}$ for 
the model parameters $q=0.1$, $c_{\rm t}^2=0.5$, 
$\lambda_1=10$, $\lambda_2=0.5$, and $V_2/V_1=10^{-6}$. 
We choose the initial values of $\hat{x}_1$ and $\hat{x}_2$ 
to coincide with those of the scaling fixed point (d) 
with $w=0$ and $q_1=q_2=0.1$. 
The initial values of perturbations are chosen to be 
$\hat{\zeta}=6.8407 \times 10^{-6}$, 
$\hat{\chi}=-2.7513 \times 10^{-6}$, 
$\hat{V}_m=3.5537 \times 10^{-9}$, 
$\hat{\delta}=2.5618 \times 10^{-3}$, and 
$\hat{{\cal K}}=30$ at $\hat{a}=2.0488 \times 10^{-3}$, under which 
the special solution (\ref{Vmspe}) is the dominant 
contribution to $\hat{V}_m$.}
\end{figure}
\begin{figure}
\includegraphics[height=3.2in,width=3.3in]{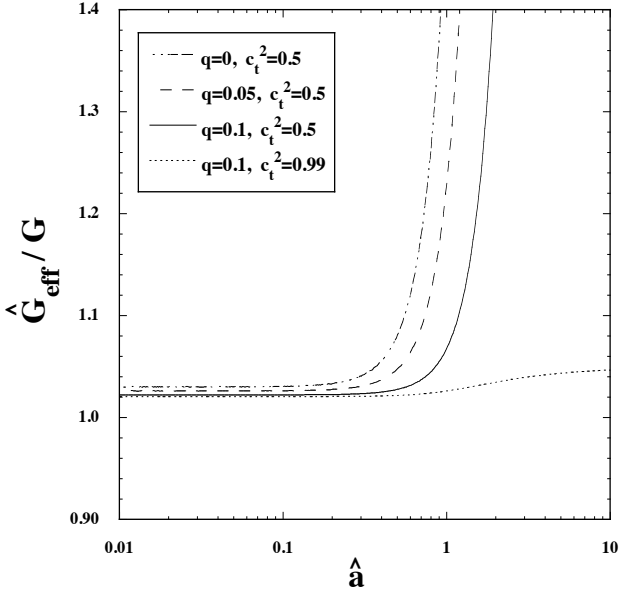}
\caption{\label{Gefffig}
Evolution of the effective gravitational coupling $\hat{G}_{\rm eff}$ 
normalized by the gravitational constant $G$ in the EF 
for the four cases: 
(i) $q=0$, $c_{\rm t}^2=0.5$, 
(ii) $q=0.05$, $c_{\rm t}^2=0.5$, 
(iii) $q=0.1$, $c_{\rm t}^2=0.5$, and 
(iv) $q=0.1$, $c_{\rm t}^2=0.99$, 
with the model parameters $\lambda_1=10$, $\lambda_2=0.5$, 
and $V_2/V_1=10^{-6}$. 
The initial conditions are chosen in the same way 
as those explained in the caption 
of Fig.~\ref{zetafig}.}
\end{figure}

In Fig.~\ref{zetafig} we plot the evolution of perturbations 
$\hat{\zeta}$, $\hat{\chi}$, $\hat{V}_m$, 
$\hat{\delta N}/\hat{\bar N}$, and $-\hat{\Psi}~(=\hat{\Phi})$ 
for the model parameters $q=0.1$, $c_{\rm t}^2=0.5$, $\lambda_1=10$, 
$\lambda_2=0.5$, and $V_2/V_1=10^{-6}$ 
(i.e., $\mu_1=9.6$, $\mu_2=0.1$, $\hat{V}_2/\hat{V}_1=10^{-6}$). 
The background initial conditions are chosen to start from 
the scaling matter fixed point (d), i.e., 
$\hat{x}_1=\sqrt{6}/[2(\mu_1+q)]$ and 
$\hat{x}_2=\sqrt{(3+2q^2+2q\mu_1)/[2(\mu_1+q)^2]}$.
We choose the initial value of $\hat{V}_m$ close to the 
special solution (\ref{Vmspe}) with $\hat{\zeta}' \simeq 0$ and 
$\hat{\chi}' \simeq 0$ for the normalized 
wave number $\hat{{\cal K}}=30$.

All the perturbations shown in Fig.~\ref{zetafig} stay nearly constant 
during the scaling matter era. We also confirmed that the analytic 
formulas (\ref{perana1})-(\ref{PhiGR}) are in good agreement 
with numerically integrated solutions in the scaling regime.
For sub-horizon perturbations ($\hat{{\cal K}} \gg 1$), the choice of 
different wave numbers $k$ only modifies the amplitudes of 
perturbations $\hat{\chi}$, $\hat{\zeta}$, $\hat{V}_m^{(s)}$, 
and $\hat{\delta N}/\hat{\bar{N}}$.
There is a simple relation $\hat{\delta N}/\hat{\bar N}
=-(3\hat{\Omega}_m/2)\hat{V}_m^{(s)}$ 
from Eqs.~(\ref{perana3}) and (\ref{perana4}). 
The perturbations $\hat{\delta N}/\hat{\bar N}$ and $\hat{V}_m^{(s)}$
are suppressed relative to $\hat{\chi}$ and $\hat{\zeta}$ 
because of the conditions $\hat{x}_1 \ll 1$ and $q \ll 1$.
The Universe finally enters the epoch of cosmic 
acceleration driven by the fixed point (c) with $\mu_2=\lambda_2-4q$.
As we see in Fig.~\ref{zetafig}, the perturbations start to vary 
after the onset of cosmic acceleration.

Numerically, we also compute the gravitational potential $\hat{\Psi}_g$ 
and the resulting effective gravitational coupling $\hat{G}_{\rm eff}$ from 
the definition (\ref{PoEin}).
In Fig.~\ref{Gefffig} the evolution of $\hat{G}_{\rm eff}/G$ is plotted 
for four different values of $q$ and $c_{\rm t}^2$. 
We confirmed that the analytic 
estimation (\ref{Geffex}) of $\hat{G}_{\rm eff}$ is in good agreement 
with the numerical result in the scaling matter regime.
The growth of $\hat{G}_{\rm eff}$ starts to occur 
after the scaling matter era.

If we compare the cases (ii) and (iii) with the case (i) 
in Fig.~\ref{Gefffig}, we find that the existence of 
coupling $q$ leads to the value of 
$\hat{G}_{\rm eff}/G$ smaller than that for $q=0$. 
Provided $c_{\rm t}^2$ is not close to 1, 
the analytic formula (\ref{Geffex}) implies that 
$\hat{G}_{\rm eff}/G$ approaches 1 in the limit where 
$q \gg \hat{x}_1$ and $\hat{x}_1 \ll 1$. 
Hence, for a given value of $c_{\rm t}^2$ different from 1, 
$\hat{G}_{\rm eff}/G$ tends to decrease with increasing $q$. 
For larger $q$, the variation of $\hat{G}_{\rm eff}/G$ occurs 
at a later cosmological epoch.

In the limit that $c_{\rm t}^2 \to 1$, Eq.~(\ref{Geffex}) reduces to 
the value $\hat{G}_{\rm eff}/G \to 1+2q^2$. 
The case (iv) in Fig.~\ref{Gefffig} is close to such an example, 
in which case the variation of $\hat{G}_{\rm eff}/G$ is small even 
after the onset of cosmic acceleration.
This property can be clearly distinguished from the model 
with $q=0$ and $c_{\rm t}^2 \neq 1$. 

If we choose initial conditions where $\hat{V}_m$ is not close to 
the special solution $\hat{V}_m^{(s)}$, the homogenous solution 
$\hat{V}_m^{(h)}$ gives rise to the oscillation of $V_m$.
This oscillation continues by the time when the amplitude of 
$\hat{V}_m^{(h)}$ decreases sufficiently relative to 
that of $\hat{V}_m^{(s)}$.
This situation is analogous to what was found
for the case $q=0$ \cite{DKT}.

\subsubsection{Jordan frame}

The evolution of perturbations in the EF can translate to that  
in the JF by using the correspondence
\ba
& &
\chi=\frac{c_{\rm t}^2}{1+\omega_H} \hat{\chi}\,,
\qquad \zeta=\hat{\zeta}\,,\qquad
V_m=\frac{\hat{V}_m}{1+\omega_H}\,,\nonumber \\
& &
\delta N=\frac{\hat{\delta N}}{\hat{\bar N}}\,,
\qquad 
\delta=\hat{\delta}\,,\qquad
{\cal K}=\hat{\cal K} \frac{1+\omega_H}{c_{\rm t}}\,,
\label{trasn1}
\ea
where ${\cal K} \equiv k/(aH)$ and 
$\omega_H=-\sqrt{6}q\hat{x}_1/(1+\sqrt{6}q\hat{x}_1)$.
The gauge-invariant matter perturbation $\delta_m$ is related to 
$\hat{\delta}_m$, as
\be
\delta_m=\hat{\delta}_m+\frac{3\omega_H}{1+\omega_H}
\hat{V}_m\,.
\label{trasn2}
\ee
At the background level, we also have the relations (\ref{backcore}) 
and (\ref{epcore}). Using the analytic 
solutions (\ref{perana1})-(\ref{perana4}) in the EF during the scaling 
matter era, the perturbations $\chi$, $\zeta$, $V_m$, and 
$\delta N$ in the JF can be expressed in terms of $\delta_m$, 
${\cal K}$, $\Omega_m$, $c_{\rm t}^2$, $x_1$, and $\epsilon_H$.

The gravitational potentials $\Psi$ and $\Phi$ are known from 
$\hat{\Psi}$, $\hat{\Phi}$, and $\hat{\chi}$ by using 
Eqs.~(\ref{Psitra}) and (\ref{Phitra}). 
Substituting the solutions (\ref{perana1}) and (\ref{PhiGR}) 
into Eqs.~(\ref{Psitra}) and (\ref{Phitra})
in the scaling matter regime and employing the approximation 
$\hat{\delta}_m \simeq \delta_m$ for the perturbations deep 
inside the Hubble radius, it follows that 
\begin{widetext}
\ba
\Psi &\simeq& 
-\frac{3[(1-c_{\rm t}^2)\epsilon_H (1+2\omega_H)+3x_1^2+\omega_H^2]}
{2{\cal K}^2[(1-c_{\rm t}^2)
(\epsilon_H+\omega_H)(1+\omega_H)
+3c_{\rm t}^2x_1^2]}\Omega_m\delta_m\,,\label{PsiJes}\\
\Phi &\simeq& 
\frac{3[c_{\rm t}^4(1+2\omega_H)+(1+\omega_H)
\{ 1+\epsilon_H+2\omega_H-c_{\rm t}^2 (2+\epsilon_H
+3\omega_H)\}+3c_{\rm t}^2 x_1^2]}{2c_{\rm t}^2
{\cal K}^2[(1-c_{\rm t}^2)
(\epsilon_H+\omega_H)(1+\omega_H)
+3c_{\rm t}^2x_1^2 ]}\Omega_m\delta_m\,,\label{PhiJes}\\
\eta &\simeq& 1+\frac{[(1-c_{\rm t}^2)(1+\epsilon_H)+2\omega_H]
[1-c_{\rm t}^2+\omega_H(1-2c_{\rm t}^2)]}
{c_{\rm t}^2 [(1-c_{\rm t}^2)\epsilon_H 
(1+2\omega_H)+3x_1^2+\omega_H^2]}\,,
\label{etaes}\\
G_{\rm eff} &\simeq& 
\frac{(1-c_{\rm t}^2)\epsilon_H (1+2\omega_H)+3x_1^2+\omega_H^2}
{(1-c_{\rm t}^2)
(\epsilon_H+\omega_H)(1+\omega_H)
+3c_{\rm t}^2x_1^2}\frac{G}{F_1}\,,
\label{GeffJes}
\ea
\end{widetext}
where $\omega_H=-\sqrt{6}qx_1$ and 
$x_1=\sqrt{6}/(2\lambda_1) \gg |\omega_H|$ for $q \ll 1$. 
The perturbation $\hat{\chi}$ induces the anisotropic stress 
in the JF, so the anisotropy parameter $\eta$ is 
different from 1.

In the limit $c_{\rm t}^2 \to 1$, the formulas 
(\ref{PsiJes})-(\ref{GeffJes}) give $\Psi \simeq -3(1+2q^2)
\Omega_m \delta_m/(2{\cal K}^2)$, 
$\Phi \simeq 3(1-2q^2)\Omega_m \delta_m/(2{\cal K}^2)$, and 
\ba
\eta &\simeq& \frac{1-2q^2}{1+2q^2}\,,
\label{eta1}\\
G_{\rm eff}
&\simeq& (1+2q^2)\frac{G}{F_1}\,.
\label{GEffJ1}
\ea

In the limit $q \to 0$ (i.e., $\omega_H \to 0$), we use the approximations 
$\epsilon_{H} \simeq 3/2$ and $\Omega_m \simeq 1$ during the scaling 
regime and eliminate the term $x_1^2$ 
on account of Eq.~(\ref{cs1}). This process leads to 
\ba
\eta &\simeq& 1+\frac{5(1-c_{\rm t}^2)(c_{\rm s}^2-c_{\rm t}^2)}
{3c_{\rm t}^2 (1+c_{\rm s}^2-c_{\rm t}^2)}\,,\label{eta2}\\
G_{\rm eff}
&\simeq&\left(1+\frac{1-c_{\rm t}^2}{c_{\rm s}^2}  
\right)G\,,
\label{GEffJ2}
\ea
which match those derived in Ref.~\cite{DKT} without referring to the EF.
Since $c_{\rm s}^2$ can be much greater than 1 for  $c_{\rm t}^2 \ll 1$, 
the parameter $\eta$ exhibits the large deviation from $1$.
In this case, $G_{\rm eff}$ is slightly larger than $G$.

\begin{figure}
\includegraphics[height=3.2in,width=3.3in]{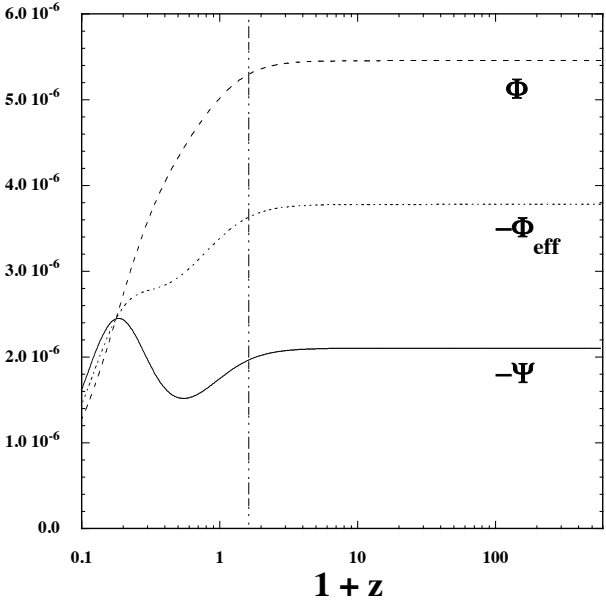}
\caption{\label{Psifig}
The gravitational potentials $-\Psi$, $\Phi$,  
and $-\Phi_{\rm eff}=(\Phi-\Psi)/2$ in the JF versus $1+z=1/a$ 
for the same model parameters and initial conditions as 
those given in Fig.~\ref{zetafig}. 
The present epoch ($z=0$) is identified by the condition 
$\Omega_m=0.3$. The vertical dot-dashed line represents 
the onset at which the cosmic acceleration ($\ddot{a}>0$) 
sets in ($z \simeq 0.63$).
Numerically, we integrate the background and perturbation 
equations in the EF and then compute $\Psi$ and $\Phi$ 
by using the transformation laws (\ref{Psitra}) and (\ref{Phitra}). }
\end{figure}

If $c_{\rm t}^2 \neq 1$ and $q \neq 0$,  the difference between 
the gravitational potentials $-\Psi$ and $\Phi$ depends on the magnitudes 
of the terms $1-c_{\rm t}^2$ and $\omega_H$. 
Provided $|1-c_{\rm t}^2| \ll \{ |\omega_H|, x_1^2 \}$, 
the parameter $\eta$ of Eq.~(\ref{etaes}) is close to the value (\ref{eta1}), 
whereas, for $|1-c_{\rm t}^2| \gg \{ |\omega_H|, x_1^2 \}$, 
$\eta$ is close to the value (\ref{eta2}).

In Fig.~\ref{Psifig} we illustrate the evolution of $-\Psi$, $\Phi$, and  
$-\Phi_{\rm eff}=(\Phi-\Psi)/2$ versus the redshift $z$ 
in the JF for $c_{\rm t}^2=0.5$, $q=0.1$, and $\lambda_1=10$.  
During the scaling matter epoch, the gravitational potentials
stay nearly constant. 
Since $|1-c_{\rm t}^2| \gg \{ |\omega_H|, x_1^2 \}$ for the model 
parameters used in the simulation of Fig.~\ref{Psifig}, we have
that $\eta \simeq 2.6$ from Eq.~(\ref{eta2}).
This is in good agreement with the numerical result of 
Fig.~\ref{Psifig} in the scaling matter regime. 
The gravitational potentials start to decrease after the onset of 
cosmic acceleration, but $-\Psi$ shows temporal growth 
in the future because of the increase of $G_{\rm eff}$ 
(see the case (iii) in Fig.~\ref{Gefffig}).
Finally, the Universe enters the attractor regime in which 
all of $-\Psi$, $\Phi$, and $-\Phi_{\rm eff}$ decrease 
in a similar way.

\begin{figure}
\includegraphics[height=3.2in,width=3.3in]{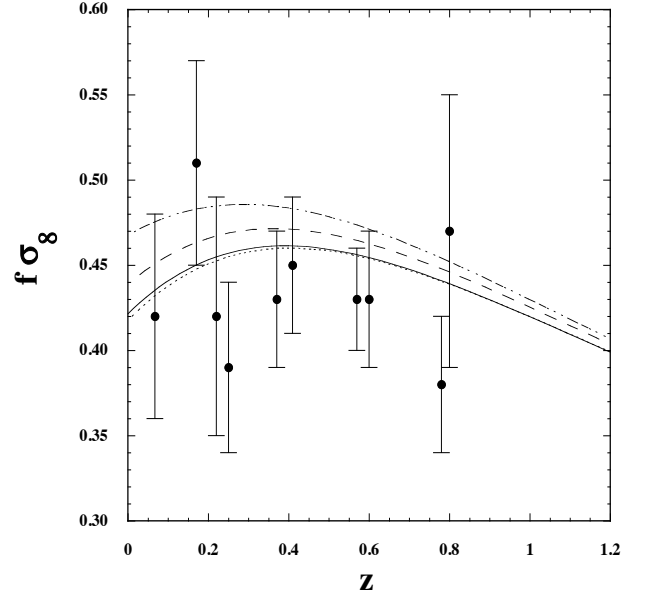}
\caption{\label{fsigmafig}
Evolution of $f(z)\sigma_8 (z)$ versus the redshift $z$ 
in the JF for $\lambda_1=10$, $\lambda_2=0.5$, and 
$V_2/V_1=10^{-6}$. 
{}From the top to the bottom, each curve corresponds to 
(i) $q=0$, $c_{\rm t}^2=0.5$, 
(ii) $q=0.05$, $c_{\rm t}^2=0.5$, 
(iii) $q=0.1$, $c_{\rm t}^2=0.5$, and 
(iv) $q=0.1$, $c_{\rm t}^2=0.99$, respectively.
The initial conditions are chosen 
in the similar way to those explained in the caption of Fig.~\ref{zetafig}. 
For a given value of $c_{\rm t}^2~(\neq 1)$, the growth rate of 
matter perturbations tends to be smaller with increasing $q$ 
due to the decrease of $G_{\rm eff}$. 
The black points with error bars correspond to the data from the recent 
observations of $f(z) \sigma_8(z)$ by 
2dFGRS \cite{2dFGRS}, 6dFGRS \cite{6dFGRS}, 
WiggleZ \cite{WiggleZ}, SDSSLRG \cite{SDSSLRG}, 
BOSSCMASS \cite{BOSSCMASS}, 
and VIPERS \cite{VIPERS} surveys.}
\end{figure}

In Fig.~\ref{fsigmafig} we plot the evolution of 
$f(z)\sigma_8 (z)$ in the low-redshift regime for four different values of $q$ and 
$c_{\rm t}^2$ (corresponding to those used in Fig.~\ref{Gefffig}), 
where $f(z)=\dot{\delta}_m/(H\delta_m)$ and $\sigma_8(z)$ 
is the rms amplitude of over-density at the comoving 
$8h^{-1}$ Mpc scale ($h$ is the normalized Hubble parameter 
$H_0 = 100\,h$ km\,${\rm sec}^{-1}\,{\rm Mpc}^{-1}$). 
This quantity is associated with the observations of red-space 
distortions in galaxy clusterings. 
Note that the growth rate $f(z)\sigma_8 (z)$ can be also measured 
by using only the peculiar motions of galaxies in the 
low redshift \cite{Johnson,Carrick}. 
The initial condition of $\delta_m$ is chosen such that its 
amplitude today is equivalent to $\sigma_8(0)=0.82$ \cite{Ade}. 
During the scaling matter era, we have numerically checked 
that $G_{\rm eff}$ is well described by Eq.~(\ref{GeffJes}). 
When $q=0$ and $c_{\rm t}^2=0.5$ we have $G_{\rm eff} \simeq 1.03G$ from 
Eq.~(\ref{GEffJ2}), whereas, for $q=0.1$ and $c_{\rm t}^2 \simeq 1$, 
$G_{\rm eff} \simeq 1.02G/F_1$ from Eq.~(\ref{GEffJ1}).

For larger $q$ with a given value of $c_{\rm t}^2~(\neq 1)$, 
the onset of growth of $G_{\rm eff}$ occurs at 
later cosmological epochs.
In fact, Fig.~\ref{fsigmafig} shows that the values of $f(z)\sigma_8 (z)$  
for $c_{\rm t}^2=0.5$ tend to be smaller with increasing $q$ 
in the low-redshift regime. 
For $c_{\rm t}^2$ close to 1, the variation of $G_{\rm eff}$ 
is small by the present epoch (see Fig.~\ref{Gefffig}).
In this case, $G_{\rm eff}$ is approximately 
given by Eq.~(\ref{GEffJ1}) even around today. 

In Fig.~\ref{fsigmafig} we also plot the recent data
of $f(z)\sigma_8(z)$ constrained by the 
redshift-space distortion (RSD) measurements. 
Using the bound on $\sigma_8(0)$ from 
the Planck data \cite{Ade}, 
the RSD data tend to favor the growth rate of 
$\delta_m$ lower than that predicted by GR \cite{weak}. 
In Fig.~\ref{fsigmafig}, such a property can be also observed 
in our model where $G_{\rm eff}$ is slightly larger than $G$.
The recent 6dF galaxy surveys using only the peculiar motions of 
galaxies provided the constraint $f(0)\sigma_8(0)=0.415 \pm 0.065$, 
which is consistent with the four cases shown in Fig.~\ref{fsigmafig}. 
It remains to see how future RSD and peculiar velocity measurements 
will pin down the error bars of $f(z)\sigma_8(z)$.

We have thus clarified the evolution of observables 
associated with the measurements of CMB, redshift-space distortions, 
and weak lensing by transforming back from the EF to the JF.
Since the EFTCAMB code \cite{EFTCAMB} for modified gravity 
theories is written in the JF, our results in this section 
can be used to place observational constraints 
on the model (\ref{LagJFex}) with the functions (\ref{AB}). 
Since the staring point of coupled 
scalar-field models (including coupled quintessence \cite{Amendola}, 
chameleons \cite{Chame}, and disformally coupled 
models \cite{Zum}-\cite{Brax}) is usually assumed to be the EF,
our analysis in the EF is also useful to constrain 
coupled dark energy models with $c_{\rm t}^2$ 
different from 1. We leave observational constraints on 
such models for a future work.
 
\section{Conclusions}
\label{consec}

In the presence of matter, we have studied 
cosmological disformal transformations in a generalized 
class of Horndeski theories (GLPV theories). 
In these theories there is one propagating scalar degree of 
freedom $\phi$ coupled to the metric $g_{\mu \nu}$ 
in the JF on the flat FLRW background. Even if matter is 
minimally coupled to gravity in the JF, the matter sector feels 
the modification of gravity through the change of 
gravitational potentials mediated by the field $\phi$.

The structure of the Lagrangian in GLPV theories, which is given by 
Eq.~(\ref{LGLPV}) in the unitary gauge, is preserved under 
the disformal transformation (\ref{diformaltra}), while the matter 
Lagrangian contains a coupling with the field $\phi$ and its derivatives 
in the transformed frame. Thus, the matter-scalar interaction becomes 
explicit after the disformal transformation.
In Sec.~\ref{disformalsec} we clarified how the energy-momentum 
tensor of matter and associated background/perturbed 
quantities are mapped under the disformal transformation.

In Sec.~\ref{transsec} we have derived the background and linear 
perturbation equations of motion in both the JF and the transformed frame.
In the transformed frame, the coupling $Q$ in Eq.~(\ref{Q})
arises for the matter continuity Eq.~(\ref{conmo}) at the background level.
The matter perturbation Eq.~(\ref{pereq4}) in the JF is also transformed to 
the more involved Eq.~(\ref{maEin}) due to the matter-scalar interaction, 
while the structure of other perturbation equations is not subject to change.

In Sec.~\ref{Einsec} we discussed the transformation from the JF to the EF 
in which the second-order action of tensor perturbations is of the same form as 
that in GR. Under the choice (\ref{choice}) of the factors 
$\Omega$ and $\Gamma$, the Bardeen potentials $\hat{\Psi}$ 
and $\hat{\Phi}$ in the EF obey the ``de-mixed'' relation (\ref{ani2}). 
If the action in the EF belongs to a class of Horndeski 
theories ($\hat{\alpha}_{\rm H}=0$), there is no anisotropic 
stress between $\hat{\Psi}$ and $\hat{\Phi}$.

The non-relativistic matter density contrast $\hat{\delta}_m$ obeys 
the differential Eq.~(\ref{delmeqEin}) in the EF, where $\hat{\Psi}_g$ 
is the effective gravitational potential given by Eq.~(\ref{Psigdef}). 
In the EF, it becomes transparent that the variations of $\Omega$ and 
${\cal C}_{\rm t}$ as well as the deviation of ${\cal C}_{\rm t}^2$ from 1
lead to the modification of gravitational interactions with matter 
perturbations. The gravitational potential $\Psi$ in the JF is simply 
related to $\hat{\Psi}_g$, as $\Psi={\cal C}_{\rm t}^2\hat{\Psi}_g$.

In Sec.~\ref{newmodelsec} we proposed a concrete model of 
dark energy in which the coupling between matter and the 
scalar degree of freedom $\phi$ is manifest after the disformal 
transformation to the EF. 
For the field potential (\ref{potential}) with $\lambda_1 \gtrsim 10$ and 
$\lambda_2 \lesssim 1$, there exist scaling solutions 
corresponding to radiation and matter eras
followed by an attractor with the cosmic acceleration. 
At the background level, the disformal transformation 
to the EF gives rise to the term $q_2$ associated with 
the function $B_4$, but we showed that the stability 
of fixed points is independent of $q_2$.
This reflects the fact that the background equations in the JF 
do not contain the function $B_4$.

We also studied the evolution of linear cosmological perturbations 
from the matter era to today for the case $q_1=q_2$.
In the EF we derived the second-order equation of the velocity 
potential $\hat{V}_m$ and identified the special solution 
$\hat{V}_m^{(s)}$ on scales deep inside the Hubble radius.
For the initial conditions satisfying 
$|\hat{V}_m^{(s)}| \gg |\hat{V}_m^{(h)}|$, we obtained 
analytic solutions of perturbations where $\hat{\zeta}$ 
and $\hat{\chi}$ stay nearly constant. On using these solutions,
we derived the effective gravitational potential $\hat{\Psi}_g$ 
of the form (\ref{PsigEF}) during the scaling matter era. 
The coupling $q$ and the deviation of $c_{\rm t}^2$ from 1 
lead to the gravitational coupling $\hat{G}_{\rm eff}$ 
modified from that in GR.

Once the evolution of perturbations is known in the EF, 
it is straightforward to transform it back to that in the JF 
by using the correspondence
(\ref{Psitra})-(\ref{Phitra}) and (\ref{trasn1})-(\ref{trasn2}).
While $\hat{\Phi}=-\hat{\Psi}$ in the EF for the Lagrangian (\ref{LagEFex}), 
the field $\hat{\chi}$ generates the anisotropic stress in the JF 
such that $\eta=-\Phi/\Psi \neq 1$.
For sub-horizon perturbations the effective gravitational coupling 
$G_{\rm eff}$ in the JF is related to $\hat{G}_{\rm eff}$ in the EF as 
$G_{\rm eff} \simeq \hat{G}_{\rm eff}/F_1$, so the growth rate of 
matter perturbations in the JF is known accordingly. 

We have analytically estimated $\eta$ and $G_{\rm eff}$ during the 
scaling matter era and confirmed that they are in good agreement 
with numerical results before the onset of cosmic acceleration.
We also numerically computed the evolution of $f(z)\sigma_8(z)$
by transforming back from the EF to the JF.
As we see in Fig.~\ref{fsigmafig}, it is possible to distinguish between 
the models with different values of $q$ and $c_{\rm t}^2$ observationally.

We have thus shown that the disformal transformation is useful for 
understanding gravitational interactions with matter mediated 
by the scalar field. After transforming to the EF, 
the background and perturbation dynamics in the JF are
readily known by using the correspondence of physical 
quantities between the two frames.
We can apply our results to observational constraints on 
dark energy models in the framework of GLPV theories. 
Moreover, our analysis in the EF is useful for constraining 
coupled dark energy models in which the starting point is
the EF Lagrangian with matter-scalar couplings.

While we focused on the cosmological set up, it is of interest to extend 
the disformal transformation to general space-time including 
the spherically symmetric background.
This should help us to understand the nature of matter-scalar couplings
in local regions of the Universe. 
In particular, the derivation of the effective gravitational coupling around a compact body (like the Sun) will be important to place constraints on theories 
beyond Horndeski from local gravity experiments.

\section*{ACKNOWLEDGEMENTS}
The author thanks Antonio De Felice, Ryotaro Kase, Kazuya Koyama, 
Federico Piazza, and Alessandra Silvestri for useful discussions. 
The author is supported by the Grant-in-Aid for Scientific Research Fund of
the JSPS (Grant No.\,24540286); MEXT Grant-in-Aid for Scientific
Research on Innovative Areas, ``Why does the Universe accelerate? -
Exhaustive study and challenge for the future'' (Grant No.\,15H05890);  
and the cooperation program between Tokyo
University of Science and CSIC.


\appendix
\section{Correspondence between the two frames}
\label{appendix} 

We show relations for the quantities connected 
through the disformal transformation. 
The background quantities are transformed as
\ba
& &
\hat{\bar{N}}=\bar{\alpha} \bar{N}\,,\qquad
\hat{H}=\frac{1}{\bar{\alpha}} \left( H+\frac{\omega}{\bar{N}} \right)\,, \nonumber \\
& & \hat{{\cal F}}=\frac{{\cal F}}{\Omega^3}\,,\qquad
\hat{\bar L}=\frac{1}{\Omega^3 \bar{\alpha}} \bar{L}\,,\nonumber \\
& & \hat{\bar{N}}\hat{L}_{,\hat{N}}=\frac{\bar{\beta}}{\Omega^3}
\biggl[\bar{L}+ \bar{N}L_{,N}-3H{\cal F} \nonumber \\
& &~~~~~~~~~~~
-\frac{1}{\bar{\alpha}\bar{\beta}} 
\left( \bar{L}-3H{\cal F}-\frac{3\omega{\cal F}}{\bar{N}}
\right) \biggr]\,.
\label{appen1}
\ea
For the quantities associated with perturbations, we have
\ba
\hspace{-0.3cm}
& &
\hat{\cal W}=\frac{\bar{\beta}}{\Omega^3}{\cal W}\,,\qquad 
\hat{L}_{,\hat{{\cal S}}}=\frac{\bar{\alpha}}
{\Omega^3}L_{,{\cal S}}\,, 
\qquad
\hat{\cal Y}=\frac{1}{\bar{\alpha} \Omega^3}{\cal Y}\,, 
\nonumber \\
\hspace{-0.3cm}
& &
\hat{\cal E}=\frac{1}{\Omega \bar{\alpha}}{\cal E}\,,\qquad
\hat{\bar{N}}\hat{\cal D}+\hat{\cal E}=
\frac{\bar{\beta}}{\Omega} 
\left(\bar{N}{\cal D}+{\cal E} \right)\,,\nonumber \\
\hspace{-0.3cm}
& &2\hat{L}_{,\hat{N}}+\hat{\bar N}L_{,\hat{N}\hat{N}}
-6\hat{H}\hat{{\cal W}}
+\frac{12\hat{L}_{,\hat{{\cal S}}}\hat{H}^2}{\hat{\bar{N}}},\nonumber \\
\hspace{-0.3cm}
& & =\frac{\bar{\beta}^2}{\Omega^3} \left( 
2L_{,N}+{\bar N}L_{,NN}-6H{\cal W}
+\frac{12L_{,{\cal S}}H^2}{\bar{N}} 
\right)+\frac{\nu \bar{\beta} \rho}{\Omega^3}\,. 
\nonumber \\
\label{appen2}
\ea

\end{document}